%% file: main.tex
\documentclass[sigconf,screen]{acmart}
\acmConference[ICSE 2024]{46th International Conference on Software Engineering}{April 2024}{Lisbon, Portugal}

\copyrightyear{2024}
\acmYear{2024}
\setcopyright{acmlicensed}\acmConference[ICSE '24]{2024 IEEE/ACM 46th International Conference on Software Engineering}{April 14--20, 2024}{Lisbon, Portugal}
\acmBooktitle{2024 IEEE/ACM 46th International Conference on Software Engineering (ICSE '24), April 14--20, 2024, Lisbon, Portugal}
\acmDOI{10.1145/3597503.3639152}
\acmISBN{979-8-4007-0217-4/24/04}

\usepackage[T1]{fontenc}
\usepackage{amsmath,amsfonts}
\usepackage{textcomp}
\usepackage{graphicx,epsfig,multirow}
\usepackage{threeparttable}
\usepackage{caption}
\usepackage{xcolor}
\usepackage{tcolorbox}
\usepackage{colortbl}
\usepackage{microtype}
\usepackage{listings}
\usepackage{float}
\usepackage{algpseudocode}
\usepackage{setspace}
\usepackage{comment}
\usepackage{fancyvrb}
\usepackage{latexsym}
\usepackage{pifont}
\usepackage{tabu}
\usepackage{booktabs}
\usepackage{xspace}
\usepackage{fancyhdr}

\usepackage{enumitem}
\usepackage{verbatim}
\usepackage{url}
\usepackage{comment}
\usepackage{tikz}
\usepackage[multiple]{footmisc}
\usepackage{pifont}
\usepackage{subcaption}
\usepackage{pifont} 
\renewcommand{\arraystretch}{1.3}

\definecolor{electricyellow}{rgb}{1.0, 1.0, 0.0}

\definecolor{light-gray}{gray}{0.8}
\definecolor{amethyst}{rgb}{0.6, 0.4, 0.8}
\definecolor{Gray}{gray}{0.87}
\definecolor{LightGray}{gray}{0.95}
\definecolor{LightCyan}{rgb}{0.88,1,1}
\definecolor{background}{rgb}{0.98,0.98,0.98}
\definecolor{codegreen}{rgb}{0,0.5,0}

\definecolor{color1}{RGB}{203, 110, 116}
\definecolor{myred}{RGB}{178, 066, 065}
\definecolor{mygreen}{RGB}{128, 154, 084}
\definecolor{color2}{gray}{0.92}
\definecolor{color3}{RGB}{236, 194, 155}
\definecolor{color4}{RGB}{243, 217, 190}
\definecolor{color5}{RGB}{239, 145, 099}
\definecolor{color6}{RGB}{255, 255, 255}

\newcommand{\tabincell}[2]{\begin{tabular}{@{}#1@{}}#2\end{tabular}}

\tikzset{
    cross/.pic = {
        \draw[rotate = 45] (-#1,0) -- (#1,0);
        \draw[rotate = 45] (0,-#1) -- (0,#1);
    }
}

\newcommand\mynuma[1]{\ifcase#1 \or \ding{172}\or \ding{173}\or
	\ding{174}\or \ding{175}\or \ding{176}\or \ding{177}%
	\or \ding{178}\or \ding{179}\or \ding{180}\or \ding{181}\else *\fi\relax}

\newcommand\mynumb[1]{\ifcase#1 \or \ding{182}\or \ding{183}\or
	\ding{184}\or \ding{185}\or \ding{186}\or \ding{187}%
	\or \ding{188}\or \ding{189}\or \ding{190}\or \ding{191}\else *\fi\relax}

\newcommand\mynumr[1]{\ifcase#1 \or \romannumeral1\or \romannumeral2\or
	\romannumeral3\or \romannumeral4\or \romannumeral5\or \romannumeral6%
	\or \romannumeral7\or \romannumeral8\or \romannumeral9\or \romannumeral10\else *\fi\relax}

\renewcommand{\smallskip}{\vspace{1mm}}

\hypersetup{
	colorlinks=true,
	linkcolor=blue,
	citecolor=blue,
	filecolor=black,
	urlcolor=black,
	pdfauthor = {},
	pdftitle = {},
	pdfkeywords = {}
}

\usepackage[strict]{changepage}

\usepackage{framed}

\definecolor{formalshade}{rgb}{0.95,0.95,1}
\definecolor{darkblue}{rgb}{0.0, 0.0, 0.55}

\newenvironment{mybullet}{\begin{list}{$\bullet$}
		{\setlength{\topsep}{0.5mm}\setlength{\itemsep}{0.5mm}
			\setlength{\parsep}{0.5mm}
			\setlength{\itemindent}{0.5mm}\setlength{\partopsep}{0.5mm}
			\setlength{\labelwidth}{15mm}
			\setlength{\leftmargin}{4mm}}}{\end{list}}

\newcommand{\conclusion}[2] {
	\begin{tcolorbox}[boxrule=1pt,left=1pt,right=1pt,top=2pt,bottom=2pt]
		\textbf{Finding:} #2
	\end{tcolorbox}
}

\newcommand{\calltoaction}[2] {
	\begin{tcolorbox}[boxrule=1pt,left=1pt,right=1pt,top=2pt,bottom=2pt]
		\textbf{Call to action:} #2
	\end{tcolorbox}
}

\makeatletter
\newcommand{\algorithmfootnote}[2][\footnotesize]{%
    \let\old@algocf@finish\@algocf@finish
    \def\@algocf@finish{\old@algocf@finish
        \leavevmode\rlap{\begin{minipage}{\linewidth}
        #1#2
        \end{minipage}}%
    }%
}
\makeatother

\newcommand{\pgftextcircled}[1]{
    \setbox0=\hbox{#1}%
    \dimen0\wd0%
    \divide\dimen0 by 2%
    \begin{tikzpicture}[baseline=(a.base)]%
        \useasboundingbox (-\the\dimen0,0pt) rectangle (\the\dimen0,1pt);
        \node[circle,draw,outer sep=0pt,inner sep=0.1ex] (a) {#1};
    \end{tikzpicture}
}

\makeatletter
\renewcommand\footnoterule{%
  \kern-3\p@
  \hrule\@width.4\columnwidth
  \kern2.6\p@}
\makeatother

\makeatletter
\def\UrlAlphabet{%
      \do\a\do\b\do\c\do\d\do\e\do\f\do\g\do\h\do\i\do\j%
      \do\k\do\l\do\m\do\n\do\o\do\p\do\q\do\r\do\s\do\t%
      \do\u\do\v\do\w\do\x\do\y\do\z\do\A\do\B\do\C\do\D%
      \do\E\do\F\do\G\do\H\do\I\do\J\do\K\do\L\do\M\do\N%
      \do\O\do\P\do\Q\do\R\do\S\do\T\do\U\do\V\do\W\do\X%
      \do\Y\do\Z}
\def\UrlDigits{\do\1\do\2\do\3\do\4\do\5\do\6\do\7\do\8\do\9\do\0}
\g@addto@macro{\UrlBreaks}{\UrlOrds}
\g@addto@macro{\UrlBreaks}{\UrlAlphabet}
\g@addto@macro{\UrlBreaks}{\UrlDigits}
\makeatother

\input{solidity-highlighting.tex}

\begin{document}

\title{Are We There Yet? Unraveling the State-of-the-Art Smart Contract Fuzzers}

\author{Shuohan Wu}
\affiliation{%
  \institution{The Hong Kong Polytechnic University}
  \city{Hong Kong}
  \country{China}
}
\email{csswu@comp.polyu.edu.hk}

\author{Zihao Li}
\affiliation{%
  \institution{The Hong Kong Polytechnic University}
  \city{Hong Kong}
  \country{China}
}
\email{cszhli@comp.polyu.edu.hk}

\author{Luyi Yan}
\affiliation{%
  \institution{The Hong Kong Polytechnic University}
  \city{Hong Kong}
  \country{China}
}
\email{cslyan@comp.polyu.edu.hk}

\author{Weimin Chen}
\affiliation{%
  \institution{The Hong Kong Polytechnic University}
  \city{Hong Kong}
  \country{China}
}
\email{cswchen@comp.polyu.edu.hk}

\author{Muhui Jiang}
\affiliation{%
  \institution{The Hong Kong Polytechnic University}
  \city{Hong Kong}
  \country{China}
}
\email{jiangmuhui@gmail.com}

\author{Chenxu Wang}
\affiliation{%
  \institution{Xi'an Jiaotong University}
  \city{Xi’an}
  \country{China}
}
\email{cxwang@mail.xjtu.edu.cn}

\author{Xiapu Luo}
\authornote{The corresponding author.}
\affiliation{%
  \institution{The Hong Kong Polytechnic University}
  \city{Hong Kong}
  \country{China}
}
\email{csxluo@comp.polyu.edu.hk}

\author{Hao Zhou}
\affiliation{%
  \institution{The Hong Kong Polytechnic University}
  \city{Hong Kong}
  \country{China}
}
\email{cshaoz@comp.polyu.edu.hk}

\input{Abstract.tex}


\maketitle
\sloppy

\input{Introduction.tex}

\input{Scope.tex}

\input{Theory.tex}

\input{Oracles.tex}

\input{bugs.tex}

\input{Evaluation}

\input{discussion.tex}

\input{Conclusion.tex}
\bibliographystyle{ACM-Reference-Format}
\bibliography{refs}

\end{document}

%% file: solidity-highlighting.tex
\definecolor{verylightgray}{rgb}{.97,.97,.97}

\lstdefinelanguage{Solidity}{
	keywords=[1]{emit, anonymous, assembly, assert, balance, break, call, callcode, case, catch, class, constant, continue, contract, debugger, default, delegatecall, delete, do, else, event, export, external, false, finally, for, function, gas, if, implements, import, in, indexed, instanceof, interface, internal, is, length, library, log0, log1, log2, log3, log4, memory, modifier, new, payable, pragma, private, protected, public, pure, push, require, return, returns, revert, selfdestruct, send, storage, struct, suicide, super, switch, then, this, throw, transfer, true, try, typeof, using, value, view, while, with, addmod, ecrecover, keccak256, mulmod, ripemd160, sha256, sha3}, 
	keywordstyle=[1]\color{blue}\bfseries,
	keywords=[2]{address, bool, byte, bytes, bytes1, bytes2, bytes3, bytes4, bytes5, bytes6, bytes7, bytes8, bytes9, bytes10, bytes11, bytes12, bytes13, bytes14, bytes15, bytes16, bytes17, bytes18, bytes19, bytes20, bytes21, bytes22, bytes23, bytes24, bytes25, bytes26, bytes27, bytes28, bytes29, bytes30, bytes31, bytes32, enum, int, int8, int16, int24, int32, int40, int48, int56, int64, int72, int80, int88, int96, int104, int112, int120, int128, int136, int144, int152, int160, int168, int176, int184, int192, int200, int208, int216, int224, int232, int240, int248, int256, mapping, string, uint, uint8, uint16, uint24, uint32, uint40, uint48, uint56, uint64, uint72, uint80, uint88, uint96, uint104, uint112, uint120, uint128, uint136, uint144, uint152, uint160, uint168, uint176, uint184, uint192, uint200, uint208, uint216, uint224, uint232, uint240, uint248, uint256, var, void, ether, finney, szabo, wei, days, hours, minutes, seconds, weeks, years},	
	keywordstyle=[2]\color{teal}\bfseries, 
	keywords=[3]{block, blockhash, coinbase, difficulty, gaslimit, number, timestamp, msg, data, gas, sender, sig, value, now, tx, gasprice, origin},	
	keywordstyle=[3]\color{violet}\bfseries,
	keywords=[4]{onlyOwner},
	keywordstyle=[4]\color{red}\bfseries,
	identifierstyle=\color{black},
	sensitive=false,
	comment=[l]{//},
	morecomment=[s]{/*}{*/},
	commentstyle=\color{gray}\ttfamily,
	stringstyle=\color{red}\ttfamily,
	morestring=[b]',
	morestring=[b]"
}

%% file: Abstract.tex
\begin{abstract}
Given the growing importance of smart contracts in various applications, ensuring their security and reliability is critical. Fuzzing, an effective vulnerability detection technique, has recently been widely applied to smart contracts. Despite numerous studies, a systematic investigation of smart contract fuzzing techniques remains lacking. In this paper, we fill this gap by: 1) providing a comprehensive review of current research in contract fuzzing, and 2) conducting an in-depth empirical study to evaluate state-of-the-art contract fuzzers' usability. To guarantee a fair evaluation, we employ a carefully-labeled benchmark and introduce a set of pragmatic performance metrics, evaluating fuzzers from five complementary perspectives. Based on our findings, we provide direction for the future research and development of contract fuzzers.

\end{abstract}

%% file: Introduction.tex
\section{Introduction}



%

Smart contracts have witnessed rapid growth and have been widely adopted for various financial purposes, such as trading, investing, and borrowing~\cite{RealWorldUse,li2023demystifying}. 
As of February 2023,  there are more than 44 million contracts deployed on Ethereum and the market cap of Ethereum has exceeded 210 billion USD~\cite{EthereumCap,qasse2023smart}.
However, despite their widespread adoption, many smart contracts lack a thorough security audit, making them vulnerable to potential attacks~\cite{sayeed2020smart}. 
In fact, a recent empirical study of 47,587 Ethereum smart contracts found potential vulnerabilities in a significant number of them~\cite{durieux2020empirical}. 
This raises concerns, especially for contracts that handle financial assets of significant value, as they are a prime target for attackers~\cite{chen2019tokenscope,chen2017under}.
For instance, in June 2016, attackers exploited a reentrancy bug in Ethereum's decentralized autonomous organization (DAO), resulting in the theft of 3.6 million Ether~\cite{DAOattack}. 
Given that smart contracts cannot be modified after deployment~\cite{li2020survey}, it is crucial to ensure their reliability before deployment.

Various approaches have been proposed to enhance the correctness and security of smart contracts, including formal verification~\cite{bhargavan2016formal,bai2018formal,abdellatif2018formal}, symbolic execution~\cite{luu2016making,mossberg2019manticore,he2021eosafe}, and other static analysis methods~\cite{grech2018madmax,feist2019slither,tikhomirov2018smartcheck,ma2021pluto}. However, each method faces limitations.
Formal verification relies on manual customized specification, which makes it difficult to automatically scale~\cite{zhuang2020smart}. 
Symbolic execution is hindered by path explosion issues when exploring complex contracts~\cite{torres2021confuzzius,brent2020ethainter}. 
Static analysis approaches analyze the control and data flow in the contract, but they can exhibit high false positives because they over-approximate contract behavior~\cite{anand2013compiler,chen2019large}.

Fuzzing has been successful in discovering vulnerabilities in traditional programs over the years~\cite{bekrar2011finding,li2018fuzzing,liang2018fuzzing}. 
Owing to its dynamic nature, it produces few false positives and can detect unexpected vulnerabilities without prior expert knowledge of vulnerable patterns~\cite{manes2019art}.
Since 2018, fuzzing has been extensively employed in the realm of smart contracts~\cite{jiang2018contractfuzzer}.
With the emergence of various smart contract fuzzers~\cite{lal2021blockchain}, it is essential to systematically evaluate their usability and effectiveness.
This can help pinpoint the strengths and weaknesses of each technique, guide researchers in selecting the most appropriate technique for their use cases, and inspire the development of new fuzzers based on the gained insights. 

However, to our best knowledge, no study has systematically investigated existing contract fuzzers and comprehensively evaluated their effectiveness.
Most related empirical works~\cite{praitheeshan2019security,kushwaha2022ethereum,zheng2023turn,chaliasos2023smart} focus on comparing static analysis-based smart contract auditing tools, while two recent studies~\cite{ren2021empirical,guo2022analysis} only evaluate a limited number of smart contract fuzzers (i.e., 3), which is far from comprehensive. 
Despite the fact that most studies that propose contract fuzzers experimentally validate their effectiveness,
we identify a lack of uniform performance metrics among them (e.g., they employ different criteria like instruction and branch coverage to evaluate code coverage). 
Furthermore, the benchmarks used in these studies are not unified. These issues lead to inconsistent and biased results, making it difficult to accurately compare different fuzzers.



To fill this gap, in the paper, we first conduct a systematic review of all existing smart contract fuzzing techniques. Then, we perform a comparative study of state-of-the-art smart contract fuzzers. 
In order to perform a fair and accurate evaluation, we need to meet the following requirements:

\noindent
$\bullet$
\textbf{(R1) Comprehensive performance metrics}: 
  Existing contract fuzzing studies use limited performance metrics, leading to biased or inaccurate comparisons. To avoid this, we need to design and use comprehensive performance metrics that can be consistently applied across different fuzzers. Evaluating fuzzer's performance from multiple perspectives can provide a more complete picture of its strengths and weaknesses.


\noindent
$\bullet$
  \textbf{(R2) Unified benchmark suite}:
  Prior smart contract fuzzing studies use distinct benchmarks, making it hard to compare the effectiveness of different fuzzers. 
  To ensure our reliability and reproducibility, we need to use a unified benchmark to test all fuzzers on the same set of contracts.
  This will create a level playing field for comparison, and ensure that our results are not skewed by the choice of benchmark contracts.

\noindent
$\bullet$
  \textbf{(R3) Correct labels}: While there are some available contract benchmarks, our experiments reveal that they are partially mislabeled. 
  This results in false positives (non-existent vulnerabilities flagged as vulnerabilities) and false negatives (actual vulnerabilities not detected), making it difficult to draw valid conclusions about different fuzzers' performance.
  Hence, we need to use a correctly labeled benchmark to ensure the reliability of insights and findings from our evaluation.


Based on the requirements, we build a benchmark consisting of 2,000 carefully-labeled smart contracts and design five categories of performance metrics. Using this benchmark and metrics, we conduct extensive experiments to compare 11 state-of-the-art smart contract fuzzers.
The experimental results show that state-of-the-art contract fuzzers are far from satisfactory in terms of vulnerability detection. 
We suggest that future fuzzers should consider enhancing their throughput, refining their test oracles, and optimizing the quality of initial seeds as potential areas of improvement.
To further grasp the real-world needs for contract fuzzers, we conduct surveys with 16 auditors from the industry. 
The survey result reveals that auditors favor fuzzers that provide convenience and flexibility in creating customized test oracles.
This broadens the scope of vulnerability detection, enabling auditors to identify logic-related bugs within their contracts.
Our study provides insights into the current state of smart contract fuzzing and suggests possible directions for future fuzzers.
In summary, we conduct the first systematic study of smart contract fuzzers:

\begin{itemize}[leftmargin=*,topsep=0pt]

\item {We review the current research advances in smart contract fuzzing based on related literature published in recent years.} 

\item {We create a benchmark with 2,000 carefully-labeled contracts and utilize it to perform a comprehensive evaluation of 11 state-of-the-art contract fuzzers. 
Our codebase and benchmark are released at \url{https://github.com/SE2023Test/SCFuzzers}}.

\item {We unveil problems in existing fuzzers and explore potential directions for the future design of fuzzers.}




\end{itemize}






%% file: scope.tex
\section{Background}
\label{sec_background}



%
\subsection{Smart Contract}
Smart contracts are programs running on blockchains (e.g., Ethereum~\cite{WelcomeEthereum,chen2019dataether}).
Smart contracts begin with compiling the source code into bytecode, and deploying it onto blockchain~\cite{SolidityLanguage,chen2020understanding}. Once deployed, each contract is assigned a unique address. 
Users can interact with the contract by encoding the function signature and actual parameters~\cite{chen2021sigrec,zhao2023deep} into a \underline{transaction} according to the \underline{Application Binary Interface (ABI)}. 
The ABI is JSON data generated in compilation, which describes methods to be invoked to execute smart contracts. 

\begin{figure}[!b]
\vspace{-1pt}
\small
\begin{lstlisting}[mathescape=true,language=Solidity, frame=none, basicstyle=\linespread{0.6} \fontsize{7}{9}\ttfamily]
contract Wallet{
  mapping(address => uint) uBalance;
  
  function deposit() public payable { //deposit money
    uBalance[msg.sender] += msg.value; }
  
  function withdraw(uint amount) public {
    require(uBalance[msg.sender] >= amount, "Fail"); 
    require(now > 30);
    msg.sender.call.value(amount)();  //reentrancy bug
    uBalance[msg.sender] -= amount;}}
\end{lstlisting}
\vspace{-1ex}
\caption{\small A vulnerable contract.}
\vspace{-2ex}
\label{fig:code1} 
\end{figure}

Smart contracts are stateful programs, and maintain a persistent storage for global states. 
The term \underline{"state variable"} in contracts refers to the global variable of a contract. 
For instance, as shown in Line~{2} of Figure~\ref{fig:code1}, \texttt{uBalance} variable, which records the balance of user accounts, is a state variable to be permanently stored in the contract's storage. The contract execution can depend on the state variable. In the example provided, a user can withdraw funds only if the user has a sufficient balance (Line~{8}). 
The state variable can only be altered by transactions, such as sending a transaction to invoke the \texttt{deposit()} function to add funds to the account (Line~{5}).

\subsection{Smart Contract Fuzzing}

\begin{figure}[!b]
	\centering
	\includegraphics[width=0.9\linewidth]{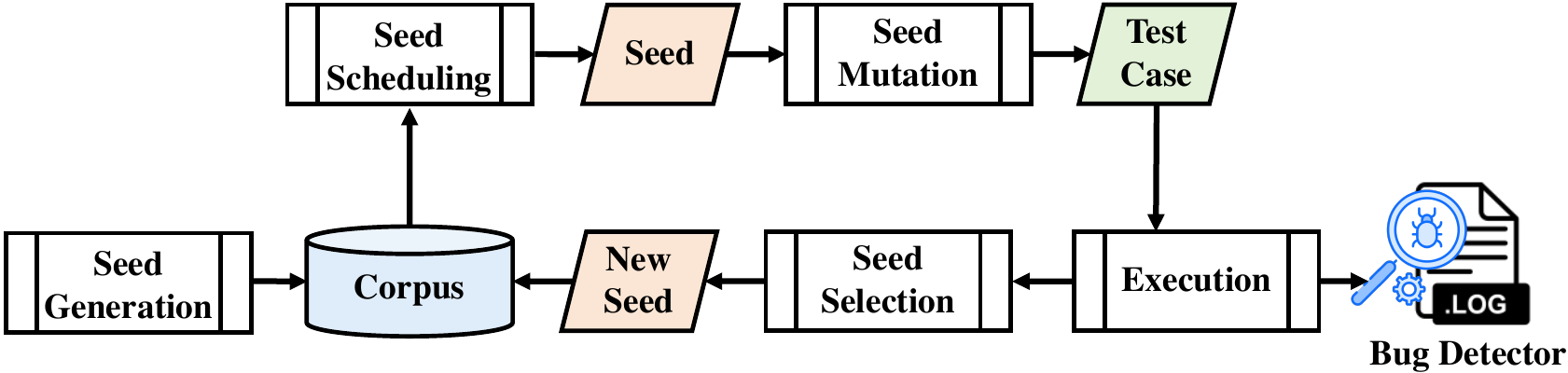}
	\caption{\small Overview of smart contract fuzzing.}
	\vspace{-2ex}
	\label{fig:contract-fuzzing-overview}
\end{figure}

Fuzzing is a widely adopted testing technique to identify defects in traditional programs~\cite{bohme2017coverage, bohme2017directed,chen2018angora,lyu2019mopt,afl,libfuzzer}. It involves generating random or unexpected data as inputs and monitoring the effects during the program's execution~\cite{zhu2022fuzzing}. 
Since 2018, fuzzing has been commonly applied to test smart contracts~\cite{jiang2018contractfuzzer}.
As shown in Figure~\ref{fig:contract-fuzzing-overview}, a typical process of smart contract fuzzing begins by constructing an initial corpus, where each seed represents a sequence of transactions. Next, the fuzzing loop selects seeds in the corpus to perform mutations, generating new inputs that are executed by the execution engine (e.g., EVM~\cite{geth}). During the execution, the engine collects runtime information, such as coverage and execution results, as feedback, which can be used to evaluate the fitness of the input. If the input achieves better fitness (e.g., higher coverage), it is retained as a seed.

In contrast to traditional fuzzing, contract fuzzing must consider \underline{transaction sequences} for executing specific functions. 
This is because the execution of a function in a transaction may depend on a contract state, which is determined by the previous transactions executed. 
As illustrated in Figure~\ref{fig:code1}, to trigger the vulnerable path (i.e., Line~{10}), 
we need first invoke \texttt{deposit()} function to deposit some funds, then call the \texttt{withdraw()} function.
Another difference is that most vulnerabilities in smart contracts are related to business logic and will not crash the program (i.e., EVM)~\cite{wang2019vultron}. 
To detect vulnerabilities, contract fuzzers typically implement test oracles within the execution engine (e.g., EVM). \underline{Test oracles} use pre-defined vulnerability patterns to match contract runtime behaviors.

\subsection{Literature Search and Scope}
This study primarily focuses on \underline{\textbf{Ethereum}} smart contract fuzzers, given Ethereum's popularity as a go-to platform for developing and deploying contracts and the abundance of research centered on it.
To identify relevant literature for our systematic survey, we set up the keywords as "smart contract fuzzing", and searched for related work published from 2016 to 2023 through seven academic databases, i.e., IEEE Explore, ACM Digital Library, Google Scholar, Springer Link, Web of Science, DBLP Bibliography, and EI Compendex. These databases maintain papers published in top conferences and journals in the field of computer science and engineering~\cite{HUTian-Yuan}.
After the initial search, we retrieved a total of 75 papers. We then read the abstract of each paper to filter out  those that do not propose a smart contract fuzzer.
For example, we removed papers focusing on theoretical aspects of contract fuzzing.
Eventually, we identified 28 papers meeting our inclusion criteria, with five focusing on other three platforms, and the remaining 23 on Ethereum. They are summarized and presented in Table~\ref{table:performance2}.


 

%% file: Theory.tex
\section{Smart Contract Fuzzing Approach}

Researchers employ various techniques to build effective fuzzers for triggering vulnerable code, including symbolic execution~\cite{he2019learning,torres2021confuzzius, choi2021smartian}, machine learning~\cite{he2019learning, zhou2021smartgift,xue2022xfuzz,liao2019soliaudit,su2022effectively}, and static/dynamic data dependency analysis~\cite{wustholz2020targeted,wustholz2020harvey,nguyen2020sfuzz,wang2020oracle,zhang2020ethploit,choi2021smartian}.
This section offers a comprehensive review of the critical components in contract fuzzing approaches. Besides, we summarize the techniques employed by different fuzzers in Table~\ref{table:performance2}, providing a clear overview of their methodologies and facilitating a thorough understanding of the current landscape.

\input{tb.tex}

\subsection{Seed Generation}
As highlighted by Herrera et al.~\cite{herrera2021seed}, constructing the initial seed is a critical step that greatly impacts fuzzing effectiveness, as a good seed can generate potential mutations to reach deeper paths and uncover vulnerabilities.
In smart contract fuzzing, seed generation involves two primary components: generating  transaction arguments and determining transaction sequences.

\noindent$\bullet$ 
\emph{\textbf{Transaction arguments generation.}}
When generating transaction arguments, 
fuzzers~\cite{pani2022smart, wesley2022verifying, liu2018reguard} built on general fuzzers (e.g., AFL~\cite{afl}) treat each argument as a byte stream and generate random bytes as inputs. Hence, they often require additional time to hit argument types (e.g., \texttt{string}). 
In contrast, Type-aware fuzzers (e.g., echidna~\cite{echidna})
extract parameter types from contracts' ABI specification, enabling them to generate values accordingly. For fixed-length arguments (e.g., \texttt{uint256}), they produce random values based on the argument's type. For non-fixed-length arguments (e.g., \texttt{string}), they typically generate a corresponding number of random elements with a random length.


In addition to randomly generating inputs, some fuzzers~\cite{jiang2018contractfuzzer, zhou2021smartgift} also utilize inputs from real-world transactions, as these inputs are more likely to trigger behaviors that mirror actual use cases.
Moreover,
ILF~\cite{he2019learning} utilizes values obtained from symbolic execution as its arguments, while ETHPLOIT~\cite{zhang2020ethploit} gathers inputs and outputs of complex functions (e.g., cryptography functions), along with magic numbers in function bodies. 
The rationale behind these approaches is that such constants may have the potential to satisfy certain branch conditions.
However, inputs from unrelated contracts may not be effective in satisfying branch conditions in the target contract. 
To address this issue, Confuzzius~\cite{torres2021confuzzius} and Beak~\cite{zhang2022beak} leverage symbolic analysis to determine input values when fuzzing stalls in certain branches. Similarly, Smartian~\cite{choi2021smartian} leverages concolic testing, which combines concrete and symbolic execution to generate arguments that satisfy certain constraints.

\noindent$\bullet$ 
\emph{\textbf{Transaction Sequence Generation.}}
As previously mentioned, the execution of smart contracts often depends on specific states, which can only be achieved through a sequence of transactions. Therefore, it is crucial to create meaningful transaction sequences in seed~\cite{wu2022review}. 
Many fuzzers~\cite{grieco2020echidna, jiang2018contractfuzzer,wustholz2020harvey,wesley2022verifying,wang2020oracle,pani2022smart,zhou2022antfuzzer} simply generate random transaction sequences as inputs. To ensure all desired functions are tested, when generating sequences randomly, a common practice is to enumerate these functions in initial seeds, as done by SynTest-Solidity~\cite{olsthoorn2022syntest}, ConFuzzius~\cite{torres2021confuzzius}, and sFuzz~\cite{nguyen2020sfuzz}. Although the random strategy can produce diverse sequences and may yield unexpected results, it is not effective in finding critical sequences due to the vast search space of all possible sequences.

Several fuzzers employ machine learning (ML) techniques to generate high-quality transaction sequences, e.g., RLF~\cite{su2022effectively}, xFuzz~\cite{xue2022xfuzz}, and ILF~\cite{he2019learning}.
ILF learns sequences from coverage-guided symbolic execution, inheriting the ability to produce high-coverage inputs with reduced overhead compared to pure symbolic execution~\cite{SurveySymExec-CSUR18}. xFuzz utilizes an ML model trained on contracts, where vulnerable functions are labeled using static analyzers (e.g., Slither~\cite{feist2019slither}). The trained model retains suspicious functions, effectively reducing the search space by excluding benign functions.
RLF categorizes contract functions into clusters based on the functionalities they provide, simplifying the function space. To generate meaningful transaction sequences, it applies deep Q-learning~\cite{bottinger2018deep}, leveraging both previous experiences and current contract state to determine the appropriate cluster from which the next function is randomly selected.
Although these fuzzers may generate effective sequences, their results can be non-deterministic due to the challenges of generalizing to unseen contracts~\cite{choi2021smartian}.

Moreover, some fuzzers incorporate data flow analysis to predict feasible transaction sequences. 
For instance, ETHPLOIT employs static taint analysis. It labels the read state variables and function arguments as taint sources, while written state variables and external calls as taint sinks. When generating transaction sequences, it selects candidate functions that can extend the taint propagation. This approach allows it to generate the seeds that reflect the data flow of real-world attack scenarios, potentially exploitable by attackers. 
Similarly, IR-Fuzz~\cite{liu2023rethinking} and Beak examine read-after-write (RAW) dependencies of global variables between functions through static analysis, while ConFuzzius tracks dynamic data flow on state variables to identify RAW dependencies for arranging transaction orders.
Besides RAW dependencies, Smartian uncovers use-def relationships of state variables using static analysis, helping explore relevant execution paths in contracts. 

\subsection{Seed Scheduling}
Effective seed scheduling is essential for smart contract fuzzing, as it determines which inputs will be used to guide the exploration of the contract's state space. 
To assess the quality of seeds, fitness metrics such as code coverage are employed. 
Seeds with higher fitness values are deemed more valuable, warranting a higher allocation of fuzzing budget. 
Next, we will delve into the different fitness metrics employed by smart contract fuzzers to optimize seed scheduling.


\noindent$\bullet$ 
\emph{\textbf{Fitness by Coverage.}}
Code coverage has long been a cornerstone metric in traditional fuzzing, and 
many smart contract fuzzers~\cite{wang2022etherfuzz,pani2022smart, wesley2022verifying, liu2018reguard,su2022effectively,choi2021smartian,nguyen2020sfuzz,xue2022xfuzz,wang2020oracle} have adopted it to prioritize seeds that can potentially uncover new code paths. Typically, these fuzzers calculate code coverage based on executed instructions and covered basic blocks, while some~\cite{choi2021smartian, torres2021confuzzius} adopt a more fine-grained approach by calculating the covered branches.
Covered branches provide a deeper understanding of a contract's behavior by examining its execution paths~\cite{branchcoverage}.
In addition to code coverage, Smartian~\cite{choi2021smartian} also incorporates data-flow coverage. This metric evaluates the dynamic data flows between state variables during the fuzzing process, providing insight into how effectively the seed exercises the contract's state variables. 
Ityfuzz~\cite{shou2023ityfuzz} tracks the value of state variables, identifying them as "interesting" if they are written with a previously unseen value. 
Both Smartian and ityFuzz adopt these strategies as they deem a smart contract's unique state vital for future exploration.

\noindent$\bullet$ 
\emph{\textbf{Fitness by Distance.}}
Distance is also a popular fitness metric, which  helps identify seeds that are more likely to reach unexplored areas.
Branch distance~\cite{zhang2022beak, ji2021increasing, nguyen2020sfuzz, olsthoorn2022syntest,wustholz2020harvey,liu2023rethinking, shou2023ityfuzz} and code distance~\cite{zhang2022beak} are two commonly used distance metrics in smart contract fuzzing, 
Branch distance evaluates how close a seed is to satisfy a missed branch by assessing the proximity of the seed to the branch condition.
For example, in the code below, if the value of msg.value in a seed is 5, its branch distance for the then-branch is 95 (100-5).
\vspace{-1ex}
\begin{lstlisting}[mathescape=true,language=Solidity, frame=none, basicstyle=\linespread{0.6} \fontsize{7}{9}\ttfamily]
function test() public {
  if (msg.value >= 100) { /**then*/}}
\end{lstlisting}
\vspace{-1ex}
This example highlights how branch distance provides more fine-grained feedback. When using code coverage, the seeds with msg.value = 99 and msg.value = 5 are considered equivalent. However, with branch distance, fuzzers prioritize the seed with msg.value = 99,  as it is more likely to satisfy the conditions after mutation.
Instead of branch distance, Beak~\cite{zhang2022beak} evaluates seeds using code distance, which calculates the distance of seed to uncovered code in the control-flow graph. Hence, a seed with a shorter code distance indicates that new paths can be reached more easily through mutation. 

Once distance metrics are obtained, fuzzers need to select the next seed for mutation based on distance fitness. 
Some fuzzers like IR-Fuzz~\cite{liu2023rethinking} prioritize seeds by taking the minimum distances, while others use more sophisticated methods that consider different objectives or optimization algorithms. For example, SynTest-Solidity~\cite{olsthoorn2022syntest} employs a many-objective optimization algorithm, DynaMOSA~\cite{panichella2017automated}, to promote inputs that get closer to the yet-uncovered branches and lines. 
Similarly, sFuzz~\cite{nguyen2020sfuzz} combines two complementary strategies: keeping seeds that increase code coverage and selecting the best seed for each just-missed branch with the closest distance.
Beak~\cite{zhang2022beak} sorts seeds by their distances and allocates energy using a simulated annealing algorithm~\cite{rutenbar1989simulated}.

\noindent$\bullet$ 
\emph{\textbf{Fitness by Vulnerability.}}
RLF~\cite{su2022effectively} prioritizes seeds based on their potential to uncover vulnerabilities. It counts the number of detected vulnerabilities to prioritize the seed that contains vulnerable function-call sequences. However, relying solely on vulnerability fitness might lead to local optima traps due to the sparsity of bugs. Therefore, RLF combines vulnerability fitness with code coverage  when scheduling seeds.



\noindent$\bullet$ 
\emph{\textbf{Other Fitness.}}
Smart contract fuzzers have adopted various other fitness metrics beyond those previously discussed. For example, 
Harvey~\cite{wustholz2020harvey} employs the Markov Chain-based scheduling strategy, which assigns more fuzzing budget to seeds that traverse the rare paths. The rationale is that rare paths require more resources because they are harder to reach than easily explored paths.
In ConFuzzius, the number of storage writes performed by the seed is also taken into account when calculating its fitness. The intuition is that in the subsequent crossover, a seed with more write operations may exhibit a higher probability of forming useful transaction sequences (i.e. sequences with RAW dependencies).

\subsection{Seed Mutation}
Seed mutation also plays a crucial role in fuzzing, enabling the generation of extensive test inputs by modifying existing seeds randomly or heuristically.
Smart contract fuzzers can perform seed mutation at three levels: transaction argument level, environment properties level, transaction sequence level.


\noindent$\bullet$ 
\emph{\textbf{Transaction argument mutation.}}
This level of seed mutation focuses on modifying the arguments of each transaction (i.e., the invoked function). 
Generation-based fuzzers~\cite{jiang2018contractfuzzer,echidna,liao2019soliaudit,zhou2021smartgift,he2019learning,su2022rlf,samreen2020reentrancy} have no seed mutation process as they do not use feedback to improve their seeds. 
Some fuzzers~\cite{ashraf2020gasfuzzer,nguyen2020sfuzz,pani2022smart,wang2020oracle,wang2022etherfuzz, zhou2022antfuzzer} utilize mutation operators found in general fuzzers (e.g., AFL), such as bit flips and additions, to generate new test inputs (For variable-length arguments, these fuzzers mutate them by pruning or padding bits).
However, many fuzzers mutate all arguments in each mutation, which may lose previously satisfied conditions~\cite{xu2019fuzzing,lu2022guan,rawat2017vuzzer}. For example, the function below has multiple conditional statements. 
\vspace{-1ex}
\begin{lstlisting}[mathescape=true,language=Solidity, frame=none, basicstyle=\linespread{0.6} \fontsize{7}{9}\ttfamily]
function test(int256 a, int256 b, int256 c) public {
  if (a < 5) {
    require (b > 10);
    /**bug;**/}}
\end{lstlisting}
\vspace{-1ex}
If the current input is (4,5,0), the execution gets stuck at Line~{3}. Mutating all three arguments (e.g., to (11,22,0)) causes previously satisfied conditions (e.g., Line~{2}) to be unsatisfied, preventing fuzzer from exploring deeper branches. Moreover, mutating all arguments at once makes it hard to identify which argument change led to a new path. 
To address this issue, Harvey~\cite{wustholz2020harvey} chooses to mutate only a single argument at a time. 
Similarly, Targy~\cite{ji2021increasing} and effuzz~\cite{ji2023effuzz} use taint analysis to find the arguments that are relevant to the target conditional branch and mutate only those arguments (e.g., argument \texttt{b} in example).


Traditional mutation operators lack insight into the input's semantics, resulting in syntactically correct but semantically invalid mutations~\cite{wang2019superion,rajpal2017not}.Hence, some fuzzers~\cite{wang2020oracle,zhang2020ethploit,echidna} opportunistically replace some arguments with values from a mutation pool.
The pool contains interesting or effective inputs that have been discovered during the fuzzing process, which can be reused to enhance the effectiveness of fuzzing.

\noindent$\bullet$ 
\emph{\textbf{Environment Properties Mutation.}}
The execution of smart contract can also be affected by environmental properties. For example, in Line~{9} of Figure~\ref{fig:code1}, the bug path can only be exercised when the timestamp is greater than 30. Therefore, many fuzzers mutate environmental properties to trigger more contract behaviors.
The sender address is commonly mutated by fuzzers, as contract usage rights can vary according to account permissions. Smartian~\cite{choi2021smartian} employs bit-flipping to mutate the sender address. Most fuzzers~\cite{liao2019soliaudit,nguyen2020sfuzz,rodler2023ef}, select the sender address from a predefined account set,  including roles like creator, administrator, user, and attacker.
Gas limits and call return values are also targeted for mutation by some fuzzers~\cite{wang2020oracle,zhang2020ethploit,torres2021confuzzius}. Modifying gas limits enables the exploration of out-of-gas exception-related behaviors, while changing contract call return values simulates diverse outcomes of called contracts, which helps identify unhandled exceptions within the contract under test.
Overall, ConFuzzius~\cite{torres2021confuzzius} stands out by mutating a more comprehensive set of environmental properties compared to others. This includes sender address, amounts, gas limits, timestamps,  block numbers, size of external code and contract call return values.
Similar to transaction argument mutation, the mutation of environment properties is mainly achieved by replacing them with a random value or reusing a previously observed value. Some tools such as sFuzz, treat them as byte streams and modify the individual bits.

\noindent$\bullet$ 
\emph{\textbf{Transaction sequence mutation.}}
This mutation generates more diverse transaction sequences to change contract's states, potentially unlocking branches guarded by them. Smartian defines three mutation operations: adding a random function, pruning a function, and swapping two functions. In addition,  fuzzers like ~\cite{nguyen2020sfuzz,ji2021increasing,olsthoorn2022syntest}  also employ a crossover operation to evolve transaction sequences. Crossover merges parts of two existing inputs to produce a new input, allowing the new input to inherit "good genes".

Beak and ConFuzzius append transactions that read from a specific storage slot to transactions that write to the same slot. This approach preserves read-after-write (RAW) dependencies and generates more reliable sequences.
Similarly, ContractMaster switches the order of two functions if they operate on the same state variable, potentially satisfying RAW dependencies.
Harvey proposes a unique approach. 
It first fuzzes the persistent states to find the functions whose coverage is affected by them.
Then, it prepends transactions that can modify the persistent state to those functions. 
By focusing on functions that are most sensitive to changes of contract's states, Harvey enables a more thorough exploration.
Pivoting from these methods, ityfuzz notes the high cost of re-executing previous transactions (in seed) to build up previous states. To address this, it directly snapshots the state and then explores it using random transactions, thus enhancing the efficiency.

\subsection{Engineering Optimization}
Smart contract fuzzers have made various engineering efforts to improve overall fuzzing performance. One approach is contract transformation, where fuzzers~\cite{liu2018reguard, wesley2022verifying, pani2022smart,ji2023effuzz} transform smart contract into native C++/Go code. This offers several optimizations:
First, the C++ compiler inlines opcode handlers in the produced native code, eliminating the interpreter loop and minimizing call overhead.
Second, the transformation reduces EVM stack operations, and it reduces the overhead caused by bounds and stack-overflow checking.
Furthermore, by converting the contracts to C++, fuzzers can leverage the mature ecosystem of fuzzing and program analysis tools. This not only saves engineering effort but also allows for the reuse of these well-refined and efficient tools (e.g., AFL, LibFuzzer~\cite{libfuzzer}).
However, it is important to note that, except for EF\textbackslash CF~\cite{rodler2023ef}, which pairs the translated C++ program with a C++ variant of EVM, these fuzzers do not use the EVM runtime as the execution environment. Consequently, they may not accurately reflect contract behavior, potentially leading to FPs or FNs.

Moreover, the majority of fuzzers~\cite{nguyen2020sfuzz,choi2021smartian,torres2021confuzzius} simulate only necessary components of the blockchain environment that are relevant to running smart contracts. This choice reduces the cost of expensive operations, such as mining new blocks and validating transactions. 
Taking it a step further, some fuzzers~\cite{shou2023ityfuzz,rodler2023ef}  adopt standalone EVMs, which focus solely on executing contract code and don't need to handle other tasks.
Another optimization is the shift towards offline program analysis or vulnerability detection, e.g., sFuzz~\cite{nguyen2020sfuzz}, which conducts vulnerability detection offline once every 500 test cases. Finally, some fuzzers employ more efficient programming languages like C/C++ to improve their execution speed~\cite{nguyen2020sfuzz,xue2022xfuzz, pani2022smart,ji2023effuzz}.

%% file: tb.tex
\begin{table*}[tbh]
\scriptsize
\centering
\renewcommand{\arraystretch}{1}
\caption{Summary of technical details for existing smart contract fuzzers.}
\resizebox{\linewidth}{!}{%
\begin{tabular}{|l|c|c|c|c|c|c|c|c|c|c|c|c|c|c|c|c|c|c|c|c|}
\hline
\multicolumn{3}{|c|}{\textbf{Tool}}&
\multirow{2}{*}{\textbf{G/M}} & 
\multicolumn{6}{c|}{\textbf{Arguments Generation}} 
&\multicolumn{4}{c|}{\textbf{Sequence Generation}}
 &\multicolumn{3}{c|}{\textbf{Seed Mutation}} 
& \multicolumn{3}{c|}{\textbf{Seed Scheduling}}
& \multirow{2}{*}{\textbf{SE}}

\\
\cline{1-3}
\cline{5-20}
Name & year & OC  &  & Type & Ran & Dict & Prev & ML & Sym & Ran & Dyn & Static & ML & Args & Env & Seq &  Cov & Dis & Bug &  \\ 
 \hline
 
\rowcolor{color2}ContractFuzzer~\cite{jiang2018contractfuzzer} & 2018 &\checkmark & G & \checkmark & \checkmark & \checkmark &   &   &   & \checkmark &   &   &   &   &   &   &   &   & &   \\
\hline

\rowcolor{color6}Reguard~\cite{liu2018reguard} & 2018 & & M &   & \checkmark &   &   &   &   & \checkmark &   &   &   &  \checkmark  &   &   & \checkmark &   &  & \\
\hline

\rowcolor{color2}ILF~\cite{he2019learning} & 2019 & \checkmark & G & \checkmark & \checkmark &   &   & \checkmark &   &   &   &   & \checkmark &  \checkmark  & \checkmark &   &   &   & &  \\
\hline

\rowcolor{color6}SoliAudit~\cite{liao2019soliaudit} & 2019 & \checkmark & G & \checkmark & \checkmark &   &   &   &   & \checkmark &   &   &   & \checkmark & \checkmark &   &   &   &  & \\
\hline

\rowcolor{color2}ContraMaster~\cite{wang2020oracle} & 2019 & \checkmark & M & \checkmark & \checkmark &   & \checkmark &   &   &  \checkmark &  \checkmark &   &   &  \checkmark & \checkmark & \checkmark & \checkmark &   &  & \\
\hline

\rowcolor{color6}Harvey~\cite{wustholz2020harvey} &2020 & & M &   & \checkmark &   & \checkmark &   &   & \checkmark &   &   &   & \checkmark  &   &  \checkmark & \checkmark &  \checkmark & &  \\
\hline

\rowcolor{color2}ETHPlOIT~\cite{zhang2020ethploit} & 2020 & \checkmark & M & \checkmark & \checkmark &   & \checkmark &   &   &   &   & \checkmark &   & \checkmark & \checkmark &   &  \checkmark &   &  & \\
\hline

\rowcolor{color6}sFuzz~\cite{nguyen2020sfuzz} & 2020 & \checkmark & M &\checkmark  &\checkmark  &   &   &   &   & \checkmark &   &   &   &\checkmark  &\checkmark  &   &  \checkmark &  \checkmark & &  \\
\hline

\rowcolor{color2}Echidna~\cite{grieco2020echidna} & 2020 & \checkmark & M & \checkmark & \checkmark & \checkmark  & \checkmark &   &   & \checkmark &   &   &   & \checkmark  &   &   & \checkmark &   & &  \\
\hline

\rowcolor{color6}GasFuzzer~\cite{ashraf2020gasfuzzer}& 2020 & & M & \checkmark & \checkmark & \checkmark &   &   &   & \checkmark &   &   &   &  \checkmark &   &   &   &   &   & \\
\hline

\rowcolor{color2}Targy~\cite{ji2021increasing} & 2021 & & M & \checkmark & \checkmark &   &   &   &   & \checkmark &   &   &   &   &\checkmark  &   & \checkmark  & \checkmark  &  & \\
\hline

\rowcolor{color6}Smartian~\cite{choi2021smartian} & 2021 & \checkmark & M & \checkmark & \checkmark &   &   &   &  \checkmark &   & \checkmark & \checkmark &   & \checkmark & \checkmark &\checkmark  & \checkmark &   & &  \\
\hline

\rowcolor{color2}SmartGift~\cite{zhou2021smartgift} & 2021 & \checkmark & G & \checkmark & \checkmark &   &   & \checkmark &   & \checkmark &   &   &   &   &   &   &   &   &  & \\
\hline

\rowcolor{color6}ConFuzzius~\cite{torres2021confuzzius} & 2021 & \checkmark & M & \checkmark & \checkmark & \checkmark  &  &   & \checkmark  & \checkmark & \checkmark &   &   & \checkmark & \checkmark & \checkmark & \checkmark &   &   &\\
\hline

\rowcolor{color2}Beak~\cite{zhang2022beak} & 2022 & & M & \checkmark & \checkmark  & \checkmark   &   &   & \checkmark  &   &   & \checkmark &   & \checkmark & \checkmark & \checkmark &   & \checkmark &  & \\
\hline

\rowcolor{color6}xFuzz~\cite{xue2022xfuzz} & 2022 & \checkmark & M &\checkmark  &\checkmark  &   &   &   &   &   &   &\checkmark &\checkmark  &\checkmark  &\checkmark  &   &  \checkmark &  \checkmark &  &  \\
\hline

\rowcolor{color2}EtherFuzz~\cite{wang2022etherfuzz} & 2022 & & M & \checkmark & \checkmark & \checkmark  &   &   &   &\checkmark  &   &   &   & \checkmark & \checkmark & \checkmark & \checkmark &   &   & \\
\hline


\rowcolor{color6}SynTest-S~\cite{olsthoorn2022syntest} & 2022 & \checkmark & M & \checkmark & \checkmark &   &   &   &   & \checkmark  &   &   &   & \checkmark  &   &  \checkmark &   &  \checkmark &  &  \\
\hline

\rowcolor{color2}RLF~\cite{su2022effectively} & 2022 & \checkmark & G & \checkmark & \checkmark &   &   &   &   &   &   &   &  \checkmark & \checkmark & \checkmark &   & \checkmark &   & \checkmark &\\
\hline

\rowcolor{color6} effuzz~\cite{ji2023effuzz} & 2023 &  &M & \checkmark & \checkmark & & & & &\checkmark & & & &  \checkmark & \checkmark & &  \checkmark &  \checkmark & & \\
\hline

\rowcolor{color2}IR-fuzz~\cite{liu2023rethinking} & 2023 & \checkmark & M & \checkmark & \checkmark &   &   &   &   &   &   & \checkmark &   & \checkmark & \checkmark & \checkmark &   & \checkmark &   & \\
\hline

\rowcolor{color6}EF\textbackslash CF~\cite{rodler2023ef} &2023 & \checkmark & M & \checkmark  &   & \checkmark  &   &   &   & \checkmark &   &   &   & \checkmark & \checkmark & \checkmark & \checkmark  &   & & \checkmark  \\
\hline

\rowcolor{color2}ityfuzz~\cite{shou2023ityfuzz} & 2023 & \checkmark & M & \checkmark & \checkmark & & & & &\checkmark & & & & \checkmark  & \checkmark  & \checkmark  & \checkmark & \checkmark & & \checkmark \\

\hline
\end{tabular}
}
\label{table:performance2}
\begin{tablenotes}[flushleft]\footnotesize
    \item {\scriptsize In the \textbf{Tool} column, OC: open-sourced; In the \textbf{G/M} column, G: generation-based, M: mutation-based; In the \textbf{Arguments Generation} column, Type: type-awareness, Ran: randomly generated arguments, Dict: arguments from a pre-prepared dictionary, Prev: values previously encountered or used, ML: arguments obtained through machine learning, Sym: arguments obtained via symbolic execution}
    \item {\scriptsize In the \textbf{Sequence Generation} column, Ran: randomly generated transaction sequences, Dyn: sequences generated through analyzing dynamic data flow, Static: sequences generated by analyzing static data flow, ML: sequences generated through machine learning;  In the \textbf{Seed Mutation} column, Args: mutating transaction arguments, Env: mutating environment properties, Seq: mutating transaction sequences; In the \textbf{Seed Scheduling} column, Cov: fitness by coverage, Dis: fitness by distance, Bug: fitness by vulnerability; The \textbf{SE} column indicates whether a standalone EVM is used.}
  \end{tablenotes}
\end{table*}

%% file: Oracles.tex
\section{Vulnerability Detection}

\subsection{Test Oracles}
\label{sec:testoracles}
Test oracles are critical components in detecting vulnerabilities during smart contracts fuzzing. As smart contracts do not crash in case of vulnerabilities and many vulnerabilities are related to business logic, their detection can be challenging. 
Existing work relies on test oracles to detect smart contract vulnerabilities. There are two main types of test oracles used in smart contract fuzzing: rule-based oracles and general-purpose oracles.


\noindent$\bullet$ 
\emph{\textbf{Rule-based oracle.}}
Rule-based oracles are widely used by smart contract fuzzers~\cite{jiang2018contractfuzzer,torres2021confuzzius,xue2022xfuzz,wang2020oracle,zhang2020ethploit,choi2021smartian,olsthoorn2022syntest,su2022rlf}.
These oracles use pre-defined rules and patterns to analyze the execution trace of contract (e.g., opcode sequences), to identify potential vulnerabilities. 
For example, to detect "Integer Overflow," rule-based oracles analyze the trace to look for the data flow from the result of an arithmetic overflow to the \texttt{CALL} instruction.

To implement rule-based oracles, most fuzzers hook into EVM to log necessary information, like executed opcodes, values of the stack, and program counter. 
This collected data is used to check the relevant rules.
Echidna and Harvey require developers to write the oracles into contract's source code. Therefore, their detection capability largely relies on the quality of oracles written by developers.




\noindent$\bullet$ 
\emph{\textbf{General-purpose oracle.}}
General-purpose oracles rely on high-level invariants to check the correctness of contract operations. These invariants are properties that hold true for all valid executions of the contract.
For example, ContraMaster~\cite{wang2020oracle} establishes invariants over the contract balance and the sum of user balances to check the correctness of transfer-related operations. If a transaction that transfers ether fails but the exception is not handled, an inconsistency will emerge between the recorded amount and the actual amount in contract, thus detecting the exception disorder bug.
However, Zhang et al.~\cite{zhangdemystifying} point out that many modern DeFi projects use complex business models that go beyond the proposed invariants in ContraMaster. 
For instance, many lending projects typically involve multiple assets and the total balances of a single asset can be volatile.








\begin{table}[thb]
\scriptsize
\centering
\vspace{-1pt}
\caption{Common smart contract vulnerabilities.} 
\vspace{-1pt}
\renewcommand{\arraystretch}{1.3}
\resizebox{\linewidth}{!}{
\begin{tabular}{|c|l|}
\hline 
\textbf{Bug Name} & \textbf{Description} \\ 
\hline
\tabincell{c}{Dangerous Delegatecall (\textbf{DD})} & \tabincell{l}{Contract uses delegatecall() to execute an untrusted code.} \\

\hline
\tabincell{c}{Block State\\ Dependency (\textbf{BD})} & \tabincell{l}{Contract uses Block states (e.g., timestamp, number) to decide \\ a critical operation (e.g., ether transfer).} \\

\hline
\tabincell{c}{Freezing Ether (\textbf{FE})} & \tabincell{l}{Contract has no function for sending Ether, or it allows \\unauthorized use of contract self destruction.} \\

\hline
\tabincell{c}{Ether Leak (\textbf{EL})} & \tabincell{l}{Contract allows arbitrary users to  retrieve ether from the contract.} \\

\hline
\tabincell{c}{Gasless  Send (\textbf{GS})} & \tabincell{l}{Contract mishandles out-of-gas exceptions when transferring \\the ether, the attackers may keep the untransferred assets.} \\

\hline
\tabincell{c}{Unhandled Exception (\textbf{UE})} & \tabincell{l}{Contract doesn't check for exception after calling external functions.} \\

\hline
\tabincell{c}{Reentrancy (\textbf{RE})} & \tabincell{l}{Contract doesn't update states (e.g., balance) 
before making an \\external call,the malicious callee reenters it and leads to a \\race condition on the state.} \\

\hline
\tabincell{c}{Suicidal (\textbf{SC})} & \tabincell{l}{Contract can be destroyed by the arbitrary user through \\ selfdestruct interface because of missing access controls.} \\

\hline
\tabincell{c}{ Integer Bug (\textbf{IB})} & \tabincell{l}{Integer operation exceeds the integer range, 
\\it can be harmful when modifying the contract’s state variables.} \\

\hline
\end{tabular}
}
\vspace{-3pt}
\label{table:EP_identification}
\end{table}

\subsection{Vulnerability Types}
Our study targets contract layer vulnerabilities, which are the focus of most fuzzers~\cite{jiang2018contractfuzzer,choi2021smartian}.
In this context, we identify nine common vulnerability types, which represent the range of vulnerabilities that current fuzzers (in Table~\ref{table:performance2}) can detect. 
Table~\ref{table:EP_identification} illustrates each of them.

%% file: bugs.tex
\begin{table*}[tb]
\footnotesize
\centering
\vspace{-2pt}
\caption{The number of vulnerabilities detected by fuzzers.} 
\vspace{-2pt}
\resizebox{0.91\linewidth}{!}{
\begin{threeparttable}
\begin{tabular}{|l|c|c|c|c|c|c|c|c|}
\hline
\textbf{Flaw} &
\textbf{ContractFuzzer} &
\textbf{ILF}  &
\textbf{RLF} &
\textbf{ConFuzzius} &
\textbf{sFuzz} &
\textbf{xFuzz} & 
\textbf{Smartian} &
\textbf{SmartGift}
\\
\hline

\textbf{DD} (29) &  TP:8, FP: 8 & TP:18, FP: 1 & TP:20, FP: 1 & TP: 22, FP: 0 & TP:\textbf{25}, FP: 2 & TP:\textbf{25}, FP: 2  & - & TP:6, FP: 8 \\
\hline

\textbf{BD} (317) &  TP:39, FP: 28  & TP:81, FP: 45 & TP: 93, FP: 57 & TP: 175, FP: 27 &  TP: \textbf{223}, FP: 32  & TP: 184, FP: 27 & TP: 102, FP: 9 & TP:38, FP: 31\\
\hline

\textbf{FE} (80) & TP:12, FP: 4 & TP:48, FP: 0 & TP:47, FP: 0 & TP:\textbf{48}, FP: 0 &  TP:2, FP: 20 & TP:1, FP: 14 & - & TP:12, FP: 4\\
\hline

\textbf{EL} (43) & -  &  TP:32, FP: 67 & TP:\textbf{38}, FP: 79 & TP:37, FP: 56  & - & - &  TP:26, FP: 84 & \\
\hline

\textbf{GS} (122) &  TP:17, FP: 5 & - & - & - &  TP:48, FP: 227 & TP:\textbf{55}, FP: 247 & - & TP:13, FP: 4 \\
\hline

\textbf{UE} (188) & TP: 14,FP: 10  & TP: 13,FP: 3 & TP: 13,FP: 3 & TP: \textbf{76},FP: 27  & TP: 41,FP: 19  & TP: 38,FP: 19 & TP: \textbf{76},FP: 23 & TP: 13,FP: 6\\
\hline

\textbf{RE} (121) & TP: 6, FP: 11 & TP: 48, FP: 9 & TP: \textbf{54}, FP: 11 & TP: 42, FP: 21 &  TP: 8, FP: 8 & TP: 8, FP: 7 & TP: 12, FP: 3  & TP: 6, FP: 9\\
\hline

\textbf{SC} (22) & -  & TP: 18, FP: 20 & TP: \textbf{20}, FP: 22 & TP: 10 , FP: 4 & - & - & TP: 8, FP: 0 & -\\
\hline

\textbf{IB} (581) & -  & - & - & TP: \textbf{566}, FP: 169  & TP: 129, FP: 106 & TP: 112, FP: 80 & TP: 413, FP: 182 & - \\

\hline

\end{tabular}

\begin{tablenotes}[flushleft]\scriptsize
    \item{\scriptsize  In \textbf{Flaw} column, \underline{the number (e.g., 29) indicates the count of contracts with each vulnerability type in our benchmark}. \\\textbf{-}: such bug oracle is not supported by corresponding fuzzers. \textbf{TP}: true positives, \textbf{FP}: false positives.}
  \end{tablenotes}
  
\end{threeparttable}
}

\label{table:comparisonDifferentTestingFramework}
\end{table*}

%% file: Evaluation.tex
\section{Evaluation}
\label{sec_evaluation}
In this section, we present a thorough evaluation of state-of-the-art smart contract fuzzers.
Given the distinct design of each fuzzer, it is challenging to evaluate their respective techniques and components in a fine-grained manner. 
Therefore, we turn to use widely accepted metrics to assess their performance across key dimensions.
To provide a comprehensive assessment, we first study the metrics adopted by previous traditional fuzzing~\cite{li2021unifuzz, klees2018evaluating} and smart contract fuzzing studies~\cite{torres2021confuzzius}, and then design five performance metrics that are tailored to smart contract fuzzers, ensuring our metrics encompass all dimensions assessed in existing studies. 
The five performance metrics we propose for evaluation are as follows:


\begin{itemize}[leftmargin=*,topsep=0pt]

\item {\textbf{Throughput}: measures the number of transactions a fuzzer can generate and execute per second, reflecting its speed and efficiency in generating tests cases.} 

\item {\textbf{Detected Bugs}: measures the average  number of vulnerabilities detected by the fuzzers, reflecting their ability to identify vulnerabilities in the smart contract.}

\item {\textbf{Effectiveness}: measures the speed of fuzzers in finding bugs, indicating how quickly a fuzzer can detect vulnerabilities.} 

\item {\textbf{Coverage}: measures the contract's code executed during fuzzing, indicating its thoroughness in exploring the contract.} 

\item {\textbf{Overhead}: measures the system resources consumed by the fuzzer during fuzzing, indicating its resource efficiency. This metric is instructive when users have limited resources.} 

\end{itemize}



\subsection{Experiment Setup}
\label{sec::Setup}

\noindent
\textbf{Tool Selection.}
We select the tools in Table~\ref{table:performance2} based on the availability of their source code. 
\underline{Five} open-source tools are excluded from our selection.
Echidna is excluded due to its property-based nature, which requires testers to manually write property tests within the contracts, demanding a good understanding of contract logic.
Additionally, we are unable to execute four tools in our environment. 
SoliAudit~\cite{liao2019soliaudit} does not offer instructions for execution.
IR-fuzz can not be installed using the provided script. 
SynTest-S~\cite{olsthoorn2022syntest} and ContraMaster~\cite{wang2020oracle} have runtime exceptions. 
Notably, the same exception persists for SynTest-S even after running the Docker image provided by the authors.
Despite our efforts to report these exceptions to authors via email,  
we have unfortunately not received their responses. 
Therefore, our experiments focus on the remaining \textbf{11} fuzzers (see tools in Figure~\ref{fig:efficiency}).





\noindent
\textbf{Environments.}  
All experiments were executed on a server equipped with an AMD EPYC 7H12 CPU (64 cores and 128 threads) running at 2.60 GHz, 512 GB memory, and operating on Ubuntu 16.04 LTS. 
To ensure a controlled and reproducible environment, each fuzzer was deployed and executed within an individual Docker container. 
The containers were allocated with 8 dedicated CPU cores, 16 GB RAM, and 2 GB swap space. 



\noindent
\textbf{Benchmark.}  
We assembled a benchmark of 2,000 smart contracts from previous research datasets~\cite{jiang2018contractfuzzer, choi2021smartian,torres2021confuzzius,ren2021empirical,ghaleb2020effective} and third-party repositories~\cite{ferreira2020smartbugs}, ensuring a diverse and representative sample. 
The average number of lines of code was 224 lines, while the average number of functions was 22.
To achieve a reliable and impartial evaluation, each fuzzer was subjected to \underline{20 repetitions} on the benchmark. 
\underline{Each} smart contract was subjected to \underline{30 minutes} of testing  by each fuzzer.

While building our benchmark, we observed that some contracts in previous datasets were incorrectly labeled. 
For example, the \texttt{buggy\_41} contract in SolidiFI Dataset~\cite{SolidiFI-benchmark} was labeled as "Integer Bug", but was found to have a "Block State Dependency" bug in Line~{76}. 
This issue affects not only the accuracy of our evaluation but also the validity of conclusions. 
To ensure a precise evaluation, it was necessary to relabel the contracts in our benchmark. 
First, we preserved their original labels. Then, we applied the 11 fuzzers to the benchmark. We perform manual inspection and relabel two groups of contracts: those reported with a vulnerability different from their original label by two or more fuzzers, and those whose originally labeled vulnerability type was not identified by any of fuzzers in our experiment.
This approach was informed by the work of Durieux et al.~\cite{durieux2020empirical}, which suggests aggregating the results of multiple analysis tools can produce more accurate outcomes.

We recruited three volunteers to manually inspect contracts requiring re-labeling.
Each volunteer inspected the contracts independently and documented their findings.
In cases where inconsistencies arose in the volunteers' results, they engaged in a discussion to reconcile their findings and reach a consensus.
This collaborative approach ensured that the relabeled benchmark accurately reflected the ground truth.

\subsection{Throughput}
\label{sec:exeSpeed}

\begin{figure}[!b]
	\centering
	\includegraphics[width=0.8\linewidth]{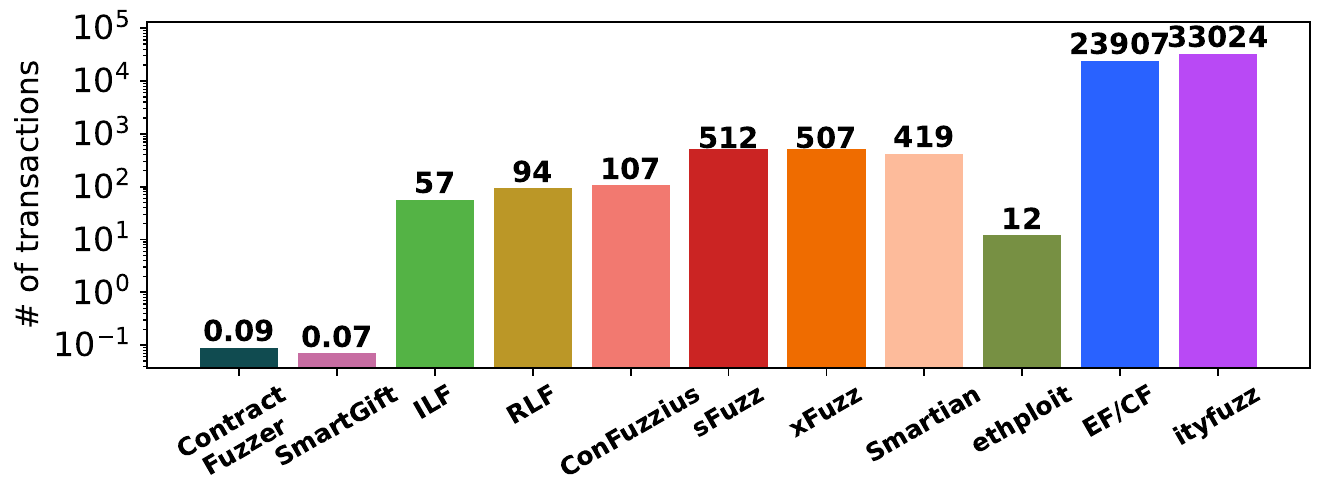}
	\caption{\small Fuzzers' throughput.}
	\label{fig:efficiency}
\end{figure}
We first evaluate the throughput of the fuzzers by measuring the average number of transactions executed per second, as shown in Figure~\ref{fig:efficiency}.
ContractFuzzer and SmartGift exhibit low throughput, generating and executing only 0.09 and 0.07 transactions per second, respectively.
This is due to their simulation of the entire blockchain network, a process that continuously appends newly mined blocks (including verification, execution of each transaction, etc.).
SmartGift uses an NLP-model for transaction arguments generation, which  further lowers its throughput.
ETHPLOIT also shows a limited throughput, primarily due to its reliance on runtime program analysis, inducing heavy overhead.
RLF, which is built upon ILF, is more efficient in test generation since it produces transaction arguments randomly, while ILF uses complex neural networks to generate the arguments. 
sFuzz and xFuzz can generate and execute tests at a more rapid pace, achieved through offline vulnerability detection (i.e., in batches, once every 500 transactions) and their avoidance of program analysis to assist fuzzing. Additionally, their integration of the well-optimized traditional fuzzer, AFL, may also help.
Both EF\textbackslash CF and ityfuzz demonstrate impressive throughput. 
This is largely due to their adoption of standalone EVMs (e.g., revm~\cite{revm}), which provide a lightweight and simplified execution environment.
Furthermore, they are coded in more efficient languages such as Rust and C++, and their superior code quality may also contribute to this performance.

\conclusion{1}{
This experiment shows that using standalone EVMs can significantly improve the fuzzers' throughput.}

\input{Problematic.tex}

\subsection{Vulnerabilities Detection Comparison}
\label{sec_bug_finding}
Table~\ref{table:comparisonDifferentTestingFramework} displays the average number of vulnerabilities detected by each tool, each running for 30 minutes per contract, over 20 repetitions.
In this experiment, \underline{we exclude \textbf{three} tools}: ETHPLOIT, ityfuzz, and EF\textbackslash CF. 
ETHPLOIT offers only vague oracles, such as "Bad Access Control", which cover several vulnerability types in Table~\ref{table:EP_identification}. 
Ityfuzz requires testers to manually write test oracles, while EF\textbackslash CF only has a general oracle that checks if the sender's balance has increased, without specifying the exact vulnerability type.
We compare the output of each tool against the labeled dataset (\S \ref{sec::Setup}) to measure their performance.
From Table~\ref{table:comparisonDifferentTestingFramework}, we observe that most fuzzers generate both high false positives and high false negatives, indicating that the state-of-the-art fuzzers are far from satisfactory in vulnerability detection.

\noindent
\emph{\textbf{False negative:}} False negatives occur due to several reasons.
Firstly, some vulnerable paths remain unexplored, protected by hard-to-satisfy branch constraints. 
Second, many fuzzers' test oracles are problematic (see Table~\ref{table:same_category}).
For instance, when detecting "Freezing Ether," ConFuzzius doesn't filter out CALL instructions introduced by swarm source contract bytecode~\cite{endcode}, causing it to miss vulnerabilities in contracts that have no ether transfer function. 
Some fuzzers' oracles are overly strict and fail to cover all vulnerable contract behaviors. One example is Smartian, which only reports "Freezing Ether" if a contract has no ether transfer instructions but contains a delegatecall that can destroy it. 
Moreover, we find some contracts cannot be executed by fuzzers due to various issues:
1) coding errors in tools (e.g., xFuzz), 2) compatibility problems with outdated Truffle (which is used to deploy contracts), 3) outdated EVM versions that don't support new Solidity instructions, and 4) outdated Solc compilers unable to process contracts written in newer Solidity versions.

\noindent
\emph{\textbf{False positive:}}
The main cause of false positives is problematic test oracles (see Table~\ref{table:same_category}).
Some test oracles are too broad, capturing non-vulnerable cases.
For instance, when detecting "Reentrancy", ILF and RLF only examine the existence of an external call followed by a storage operation in the traces, neglecting the data dependency between them. 
"Integer Under/Overflow" is often falsely reported by most fuzzers, as their oracles only syntactically check for data flow between arithmetic operations and store instructions, ignoring the actual effects on contracts (as there can be some arithmetic operations introduced by the masking operations from the compiler or compiler's optimization).
Furthermore, all these fuzzers fail to consider the context of contracts (i.e., their actual semantics), such as existing access control mechanisms in place. This may produce false positives since they might flag a behavior as vulnerable even it is not exploitable in practice. 
Additionally, lack of context prevents fuzzers from understanding contracts' higher-level intentions or purpose. 
Consequently, they struggle to differentiate between the contract's intended behavior and actual malicious one. For example, fuzzers may incorrectly label lottery contracts~\cite{atzei2017survey} as "Ether Leak".

\conclusion{1}{
Existing contract fuzzers are far from satisfactory in terms of vulnerability detection.
Their adoption of overly generalized or overly specific test oracles contributes to high false positives and negatives.
\\
\textbf{Call to action: Refine test oracles.}
To improve vulnerability detection, fuzzers should implement precise, comprehensive rules in oracles, rooted in the nature of vulnerabilities.
This involves discerning vulnerability indicators and impacts.
The incorporation of data flow analysis and machine learning (trained on vulnerability datasets) can further enhance oracles.
Furthermore, ongoing effort is needed to expand understanding of vulnerabilities, uncover new attack vectors, and advance test oracles.
}


\subsection{Speed of Vulnerability Detection}
\begin{figure}[!t]
	\centering
	\includegraphics[width=0.8\linewidth]{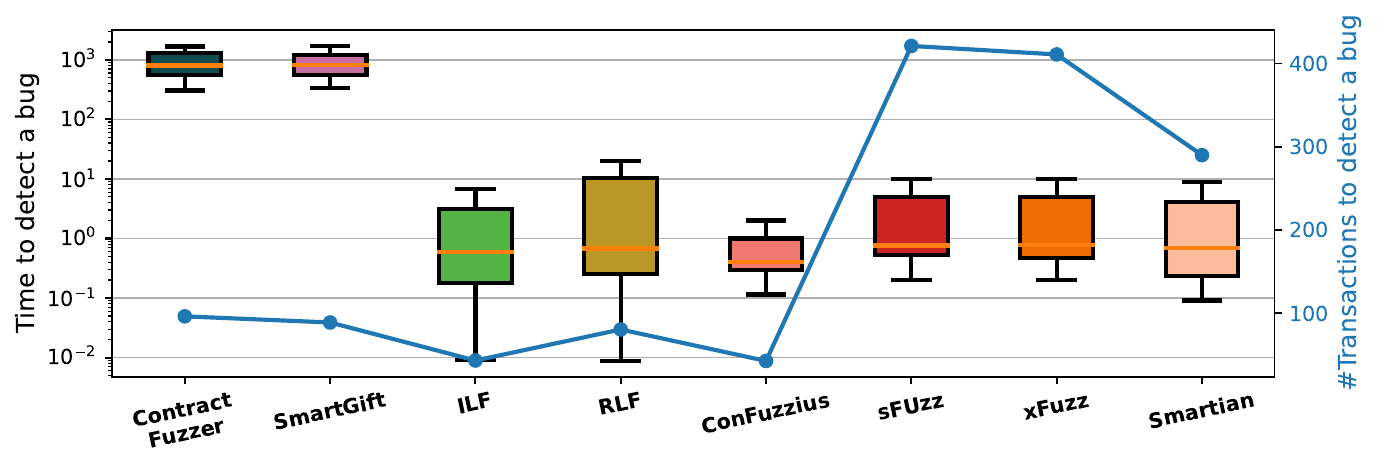}
	\vspace{-4pt}
	\caption{\small Time to find vulnerability.}
	\vspace{-2pt}
	\label{fig:timetobug}
\end{figure}

We evaluate the average time taken by each tool to detect a vulnerability (true positive). Figure~\ref{fig:timetobug} (box plot on the \underline{left} Y-axis) presents the results, for most fuzzers, they can detect a true vulnerability within 1 second. 
ConFuzzius, with its smaller upper outliers, benefits from symbolic taint analysis and data dependency analysis, expediting the discovery of vulnerable paths in certain cases.

To fairly compare fuzzing strategies of different fuzzers, irrespective of their throughput, we compute the number of transactions each fuzzer requires to find a vulnerability. 
This is achieved by multiplying the median time from the box plot in Figure~\ref{fig:timetobug} by the throughput in Figure~\ref{fig:efficiency}. 
We use the median rather than average, as it is typically more representative, reducing the impact of extreme outliers. 
The result is shown by the line graph in Figure~\ref{fig:timetobug} (on the \underline{right} Y-axis).
The last three, sFuzz, xFuzz, and Smartian, require more transactions. Because sFuzz and xFuzz batch the vulnerability detection (roughly 500 transactions, this count may vary due to sensitive instructions present).
Similarly, Smartian also experiences a detection delay. For example, in the case of "Reentrancy", the oracle only triggers to check for CALL instructions in previous transactions after the SSTORE (in current transaction) is executed.
For the first five fuzzers, ContractFuzzer and SmartGift which require notably longer times to detect a vulnerability, maintain transaction counts for actual detection comparable to other three. 
This suggests that the speed of vulnerability detection is largely dependent on fuzzers' throughput.


\vspace{-2ex}
\begin{tcolorbox}[boxrule=1pt,left=1pt,right=1pt,top=1pt,bottom=1pt]
\textbf{Finding:} In our experiment, the speed of vulnerability detection is largely influenced by fuzzers' throughput.
\\
\textbf{Call to action: Improve throughput.} 
To speed up vulnerability detection, fuzzers should take actions to increase throughput. This could be achieved by adopting lightweight, standalone EVM frameworks, conducting code optimizations, using efficient languages such as C, incorporating optimized libraries like libAFL.
\end{tcolorbox}

\vspace{-4ex}

\subsection{Code Coverage}
\begin{figure}[thb]
	\centering
	\includegraphics[width=0.88\linewidth]{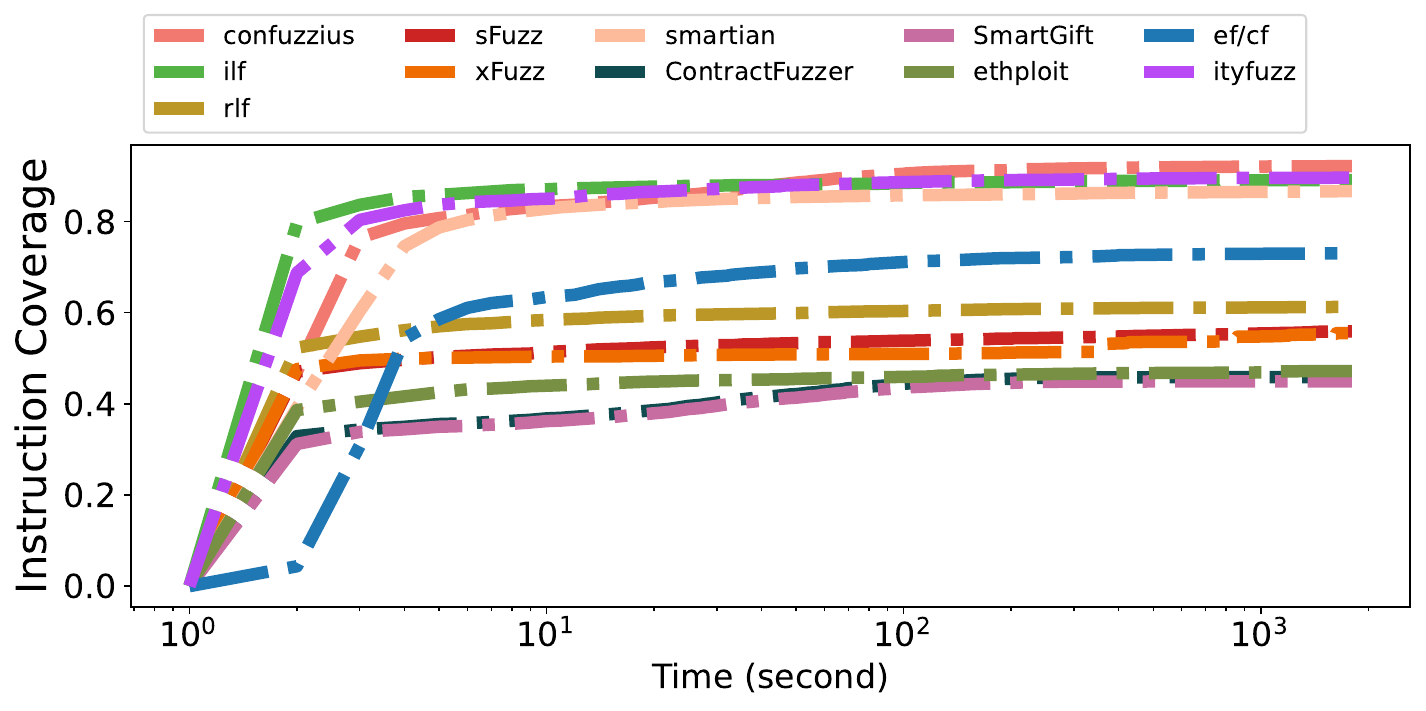}
	\vspace{-2pt}
	\caption{\small Instruction coverage.}
	\vspace{-2pt}
	\label{fig:inscoverage}
\end{figure}

We next evaluate the code coverage achieved by \textbf{11} contract fuzzers. 
To ensure a consistent comparison, we instrument the EVM uniformly to measure instruction coverage. 
Figure~\ref{fig:inscoverage} presents the results.
Notably, ityfuzz, Smartian, ILF, and ConFuzzius achieve higher code coverage compared to other tools.
This is largely due to their use of coverage-guided approaches in test case generation. 
In addition, they all mutate the environmental properties during seed mutation, which crucially simulates diverse conditions that trigger different contract behaviors.
ILF shows the fastest coverage increase in initial stages, albeit with a limited increment in later stages. This is likely due to its high-quality initial seed. It pre-trains a model from coverage-guided symbolic execution expert for generating transaction sequences and arguments.
ityfuzz also exhibits rapid initial coverage growth, due to its impressive throughput.
EF\textbackslash CF starts with a slow increase in coverage, as it begins with a colorization stage~\cite{aschermann2019redqueen}. 
In contrast, ContractFuzzer and SmartGift exhibit the lowest coverage due to two main factors. First, their throughput is markedly slow (\S \ref{sec:exeSpeed}). 
Second, as black-box fuzzers, they do not use feedback to guide test generation. 
Although SmartGift utilizes practical inputs from similar functions in the real world, its ML approach suffers from generalization issues (i.e., many test functions are not well-represented in the training set).

The code coverage in some fuzzers is lowered due to their focus only on potentially vulnerable functions during test generation. For example, sFuzz, ETHPLOIT, and RLF exclude 'View' and 'Pure' functions in contracts that do not modify state variables. 
The rationale behind this decision is that these functions are unlikely to be exploited by attackers since they have no impact on the contract states~\cite{li2022redefender}.
Similarly, xFuzz filters out benign functions using an ML model to reduce the search space of function call combinations.

In some cases, all these fuzzers struggle to achieve effective coverage. Hard constraints in complex contracts, like those relying on a 'magic' number, are difficult to satisfy. Even fuzzers incorporating symbolic execution cannot solve these constraints due to the path explosion. Additionally, reaching contract states that require long transaction sequences remains challenging. While some fuzzers analyze data flow between transactions (e.g., RAW) to arrange their sequences, they have difficulty determining transaction dependencies due to potential circular dependencies between state variables.


\conclusion{1}{
The quality of initial seeds and fuzzers' throughput are pivotal for quickly reaching high coverage.
\\
\textbf{Call to action:}\\ 
$\bullet$ 
\textbf{Enhancing initial seeds.} 
To enhance coverage effectiveness,
fuzzers should use techniques such as program analysis and machine learning to build high-quality initial seeds, which in turn, generate more effective and meaningful test inputs.
\\
$\bullet$
\textbf{Optimize
mutation scheduling.} 
Currently, all existing fuzzers randomly choose mutations from a list of mutational operators, which may waste fuzzing iterations on ineffective mutations. 
Recent research in traditional fuzzing~\cite{jauernig2022darwin, wu2022one} highlights the pivotal role mutation scheduling plays in enhancing coverage effectiveness. 
Therefore, future fuzzers should consider adopting advanced methods such as genetic algorithms or differential evolution algorithms to improve mutation scheduling.
}

\begin{figure}[thb]
    \centering
    \begin{minipage}[]{0.49\linewidth}
        \centering
        \begin{subfigure}[b]{1\textwidth}
        \centering
        \includegraphics[width=1\textwidth]{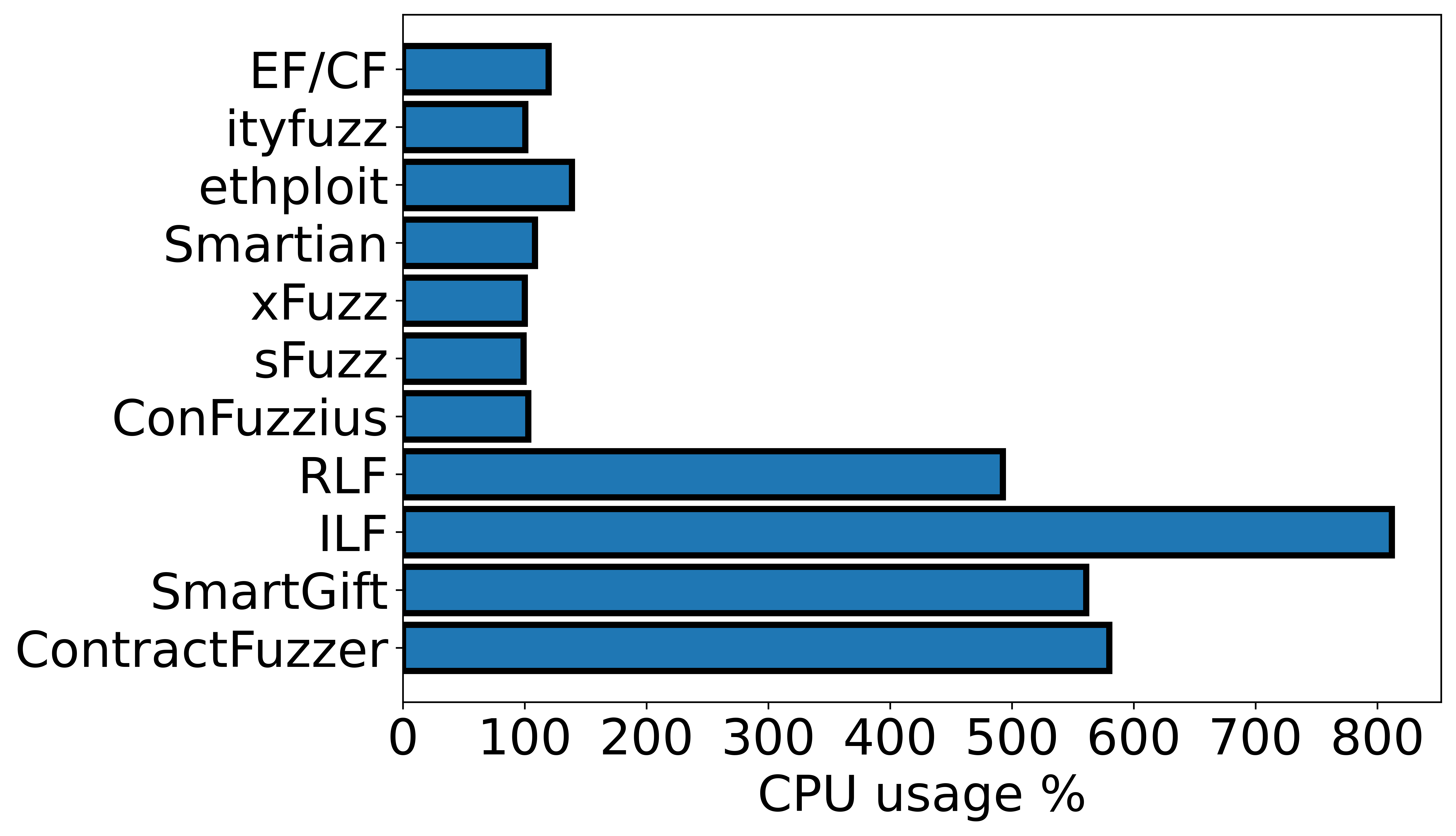}
        \vspace{-3ex}
        \caption{\small CPU Consumption.}
        \label{fig:NAVExample}
        \end{subfigure}
    \end{minipage}
    \hfill
    \begin{minipage}[]{0.49\linewidth}
        \centering
        \begin{subfigure}[b]{1\textwidth}
        \centering
        \includegraphics[width=1\textwidth]{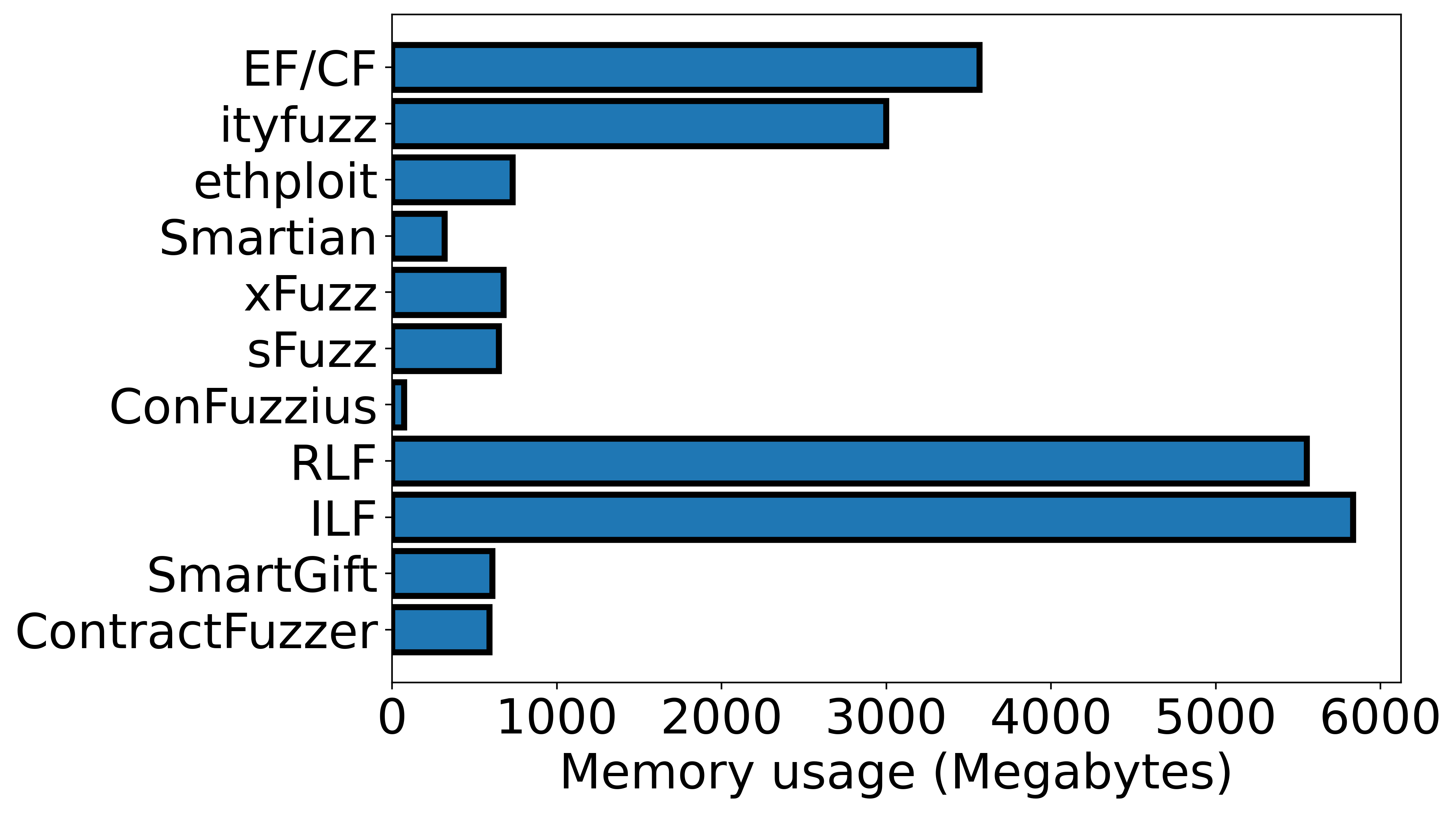}
        \vspace{-3ex}
        \caption{\small Memory Consumption.}
        \label{fig:TabExample}
        \end{subfigure}
    \end{minipage}
    \hfill
    \caption{Overhead.}
    \label{fig:overhead}
    \vspace{-1ex}
\end{figure}

\subsection{Overhead}
We next assess the overhead of each fuzzer.
We use the "docker stats" command to monitor the  resource usage in the Docker containers. Figure~\ref{fig:overhead} reports the result.
Overall, ConFuzzius exhibits the lowest overhead, as it uses PyEVM, which is a lightweight implementation of EVM.
SmartGift and ContractFuzzer exhibit high CPU usage, primarily because they simulate the entire blockchain network, which can be very CPU-intensive. 
ILF employs complex neural network models for both transaction argument and sequence generation, which requires significant computational resources. 
Since the neural network used in ILF is more complex than the DNN used in RLF, more processing power is needed.
The use of neural networks also results in both RLF and ILF consuming more memory, with an average memory consumption of 5.83 GB. This is due to the large number of parameters involved in neural networks that need to be stored in memory.

\vspace{-1ex}
\begin{tcolorbox}[boxrule=1pt,left=1pt,right=1pt,top=1pt,bottom=1pt, colback=white!20, colframe=blue!70]
\textbf{Conclusion:} ConFuzzius stands out in most key metrics. It identifies the most vulnerabilities and does so at the highest speed, achieving top-tier coverage and maintaining manageable overhead.
However, there's still room for ConFuzzius to improve. 
For example, using a standalone EVM could boost its throughput, and refining its test oracle could elevate the detection accuracy.
\end{tcolorbox}
\vspace{-3ex}






%% file: Problematic.tex
\begin{table*}[tbh]
\scriptsize
\centering
\caption{Problematic test oracles in smart contract fuzzers.} 
\resizebox{\linewidth}{!}{
\begin{tabular}
{|l|l|c|c|}
\hline
\textbf{Oracle}  &
\textbf{Issue Description} &
\textbf{Fuzzers} &
\textbf{Fault}\\
\hline
\multirow{3}{*}{\textbf{RE}} &
\tabincell{l}{Only check for the presence of external call followed by storage operation, ignoring data dependency.} &
\tabincell{l}{ILF, RLF}  & FP \\
\cline{2-4}
& \tabincell{l}{Only validate if a method can be re-entered but ignore the storage access.} &
\tabincell{l}{ContractFuzzer, SmartGift, sFuzz, xFuzz} & FP\\
\cline{2-4}
& \tabincell{l}{Taint analysis over-approximates the possible execution paths.} & \tabincell{l}{ConFuzzius} & FP\\
\cline{2-4}
& \tabincell{l}{
Check for the use of any state variables in cyclic calls, regardless of whether there are any write operations performed on them.
} & \tabincell{l}{Smartian} & FP\\
\hline


\multirow{4}{*}{\textbf{BD}} &
\tabincell{l}{Only check for ether transfer and block state related instructions (e.g., TIMESTAMP), ignoring the dependency between them.} &
\tabincell{c}{ContractFuzzer,SmartGift, sFuzz, xFuzz}  & FP \\
\cline{2-4}
& \tabincell{l}{Consider all block-related information that flows into condition statements as vulnerable.} &
\tabincell{l}{ConFuzzius} & FP\\
\cline{2-4}
& \tabincell{l}{Fail to consider indirect taint flow propagation, only look for direct taint flow.} & \tabincell{l}{ILF, RLF} & FN\\
\cline{2-4}
& \tabincell{l}{Fail to check if the block-related information affects the conditional branch.} & \tabincell{l}{Smartian} & FN\\
\hline

\multirow{2}{*}{\textbf{GS}} &
\tabincell{l}{Roughly classify all external calls with 2300 gaslimit as vulnerable, without checking the occurrence of out-of-gas exception.} &
\tabincell{l}{sFuzz, xFuzz}  & FP \\
\cline{2-4}
& \tabincell{l}{Fail to recognize that the \texttt{transfer()} automatically reverts the program state when there is not enough gas.} &
\tabincell{l}{ContractFuzzer, SmartGift} & FP\\
\hline

\multirow{1}{*}{\textbf{DD}} &
\tabincell{l}{Not check if the delegatecall parameters can be controlled by user input.} &
\tabincell{l}{ContractFuzzer, SmartGift}  & FP \\
\hline

\multirow{2}{*}{\textbf{IB}} &
\tabincell{l}{Only check the flow of arithmetic operations, ignoring their actual effects on the contract.} &
\tabincell{c}{Smartian, ConFuzzius, sFuzz, xFuzz }  & FP \\
\cline{2-4}
& \tabincell{l}{Fail to consider MUL instruction.} &
\tabincell{l}{sFuzz, xFuzz} & FN\\
\hline

\multirow{1}{*}{\textbf{EL}} &
\tabincell{l}{Fail to consider the case where the ether transfer can be reverted when the account has no balance.} &
\tabincell{l}{ConFuzzius}  & FP \\
\hline

\multirow{4}{*}{\textbf{UE}} &
\tabincell{l}{Only check for 'INVALID' in the execution trace, without considering thrown exception with a REVERT instruction.} &
\tabincell{l}{sFuzz, xFuzz}  & FN \\
\cline{2-4}
& \tabincell{l}{Only consider the case where the exception is handled immediately.} &
\tabincell{l}{ILF, RLF} & FP\\
\cline{2-4}
& \tabincell{l}{Z3~\cite{z3} inaccurately solves tainted stack items, which symbolically represent exception-related variables.} &
\tabincell{l}{ConFuzzius} & FP\\
\cline{2-4}
& \tabincell{l}{Roughly checks for all unused return values, even though not every return value  indicates a failed call.} &
\tabincell{l}{Smartian} & FP\\
\hline

\multirow{2}{*}{\textbf{FE}} &
\tabincell{l}{Fail to consider cases where a contract can receive ether but has no functions to send ether.} &
\tabincell{l}{sFuzz, xFuzz}  & FN \\
\cline{2-4}
& \tabincell{l}{Fail to consider cases where a contract with a self-destruct operation can lock the ether.} &
\tabincell{l}{ILF, RLF} & FN\\
\cline{2-4}
& \tabincell{l}{Fail to filter out CALL instructions introduced by the swarm source of bytecode, which are wrongly treated as ether transfer.} &
\tabincell{l}{ConFuzzius} & FN\\
\hline

\multirow{1}{*}{\textbf{SC}} &
\tabincell{l}{Incorrectly treats the owner account as an attacker account.} &
\tabincell{l}{ILF, RLF}  & FP \\
\cline{2-4}
\hline
\end{tabular}
}
  
\label{table:same_category}
\end{table*}

%% file: discussion.tex
\section{Industrial Perspectives}

\subsection{Fuzzing Tools Used by Auditing Companies}
Despite recent research~\cite{chaliasos2023smart,zhangdemystifying} suggesting that academic fuzzers fall short in detecting real-world contract vulnerabilities, we find that the industry continues to rely on fuzzing as an important component of contract auditing~\cite{consensys}.
To better understand the practical usage of existing smart contract fuzzers in the industry, we explore their use by various auditing companies. 
Specifically, we collect public audit reports from their respective GitHub repositories. 
The companies considered are drawn from a list of notable Blockchain Security auditing companies~\cite{audit-list}.
By scanning these reports, we identify the fuzzers employed in their auditing process. 
We also manually read the auditing methodology sections on their official websites to extend our knowledge of the employed fuzzers.
The result is reported in Table~\ref{table:auditingcompany}. 
Echidna (a product of Trail of Bits~\cite{trailofbits}) is the most favored tool among these companies.
Conversely, most of the fuzzers listed in Table~\ref{table:performance2} from academic research are not actively employed in real-world security audits. 
This highlights the gap between theoretical research advancements and the actual needs for security auditing in the industry.

\begin{table}[thb]
\centering
\vspace{-1pt}
\caption{Fuzzing tools used by different auditing companies.} 
\vspace{-1pt}
\renewcommand{\arraystretch}{1.3}
\resizebox{\linewidth}{!}{
\begin{tabular}{|c|l|}
\hline 
\textbf{Tools} & \textbf{Auditing Company} \\ 
\hline
Echidna~\cite{echidna} & \tabincell{l}{ImmuneBytes, Halborn, Trail of Bits, QuillAudits, Solidified,\\Pessimistic, ChainSafe, yAcademy, yAudit, Truscova, Zellic,\\ Zokyo, Cyfrin, ABDK.} \\

\hline
sfuzz~\cite{nguyen2020sfuzz} & \tabincell{l}{ImmuneBytes.} \\

\hline
\end{tabular}
}
\vspace{-2pt}
\label{table:auditingcompany}
\end{table}

\begin{figure}[thb]
	\centering
	\includegraphics[width=\linewidth]{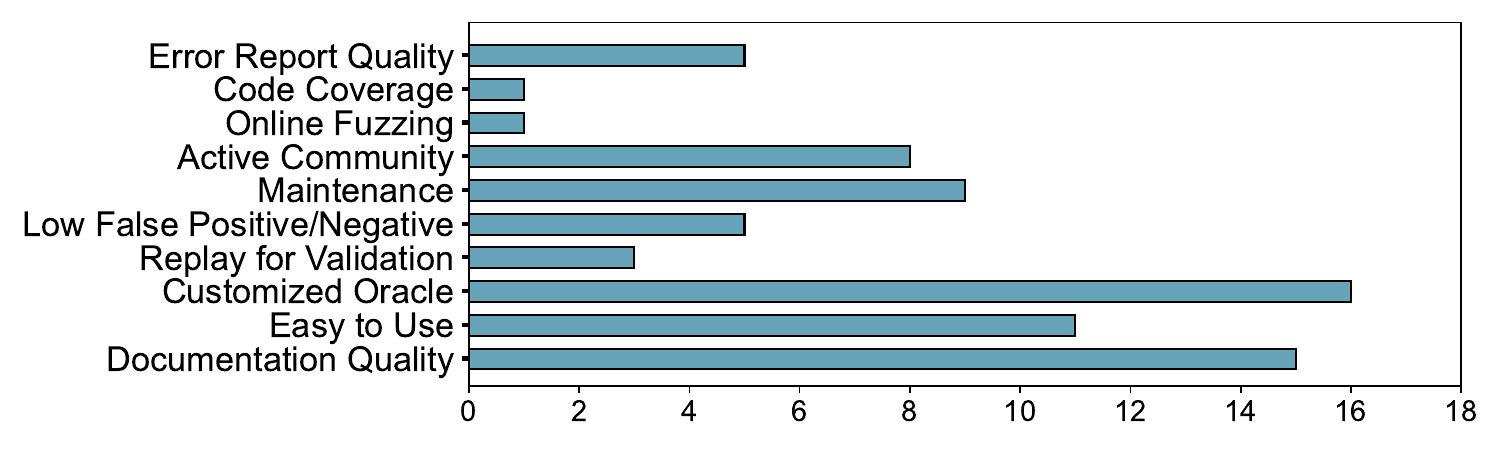}
	\vspace{-4pt}
	\caption{\small Essential factors in fuzzer selection by auditors; x-axis represents the number of participants.}
	\vspace{-2pt}
	\label{fig:surveyresult}
\end{figure}

\subsection{Survey on Auditors}
To further understand the practical needs for fuzzers, we conducted an anonymous online survey. 
We recruited 16 participants via our professional contacts within the community, who averaged 1.68 years of experience in smart contract review.
Of these, 11 were solo auditors, with the rest working in three auditing companies.
Our survey primarily questioned the participants about the fuzzers they employ during auditing, their reason for these choices, and the essential factors they consider when using fuzzers.
Due to space constraints, the complete survey questionnaire is accessible in our GitHub repository.
The survey result echoed our previous findings: 14 participants use Echidna to aid their audits, two use Foundry~\cite{foundry} which supports fuzzing in its unit tests, and one also uses ityfuzz. 

Regarding their preferences, all Echidna users praise its highly conducive support for custom oracles. As Echidna uses property-based testing~\cite{grieco2020echidna}, auditors can write invariants (constants that should perpetually hold true, see \ref{sec:testoracles}) as customized oracles.
This allows them to examine the business logic in their contracts, thereby making Echidna applicable to logic-related vulnerabilities.
Upon reviewing the fuzzers listed in Table~\ref{table:performance2}, it's evident that, besides Echidna, only ityfuzz offers substantial support for custom oracles. In contrast, the rest require developers to have a deep understanding of the source code and the ability to modify it.
In addition, three participants underscore Echidna's straightforward integration into the audit workflow.
This implies that auditors can use it with other tools in their toolkit to extract crucial information prior to and during the fuzzing campaign.
For instance, it can be readily paired with Slither~\cite{feist2019slither}, a highly recognized static analysis tool for Solidity contracts (a product from the same company). 
One participant, a former Echidna user, shifts to ityFuzz due to its support for online fuzzing  – a feature currently exclusive to it. 
This feature is important in detecting real-world vulnerabilities, given its capacity to test smart contracts in a realistic setting compared to offline fuzzing, which starts from a blank contract state. 
This participant also notes another advantage of online fuzzing - the ability to detect cross-contract vulnerabilities, as it involves interacting with other deployed contracts, a complexity that offline fuzzing fails to mimic.
As for the essential factors in selecting fuzzers, most auditors value ease of use, quality of documentation, oracle customization flexibility, and continuous code maintenance. Detailed results can be seen in Figure~\ref{fig:surveyresult}.

\calltoaction{1}{

\noindent\textbf{$\bullet$ Promoting Custom Oracle Creation.}
Fuzzers should facilitate users in creating customized test oracles in a straightforward and flexible manner. This feature allows for the capture of complex business rules, broadening the fuzzer's applicability and leading to more accurate audits. 


\noindent\textbf{$\bullet$ Support online-fuzzing.}
Fuzzers should include online fuzzing features for testing in real settings, facilitating cross-contract vulnerabilities detection
and increasing chances of discovering real exploits.
%

\noindent\textbf{$\bullet$ Emphasize Usability.} 
Fuzzers should uphold the quality of their documentation and provide clear tutorials for quick user onboarding. 
Furthermore, an intuitive interface that provides comprehensive information is also crucial. It can aid users monitoring the fuzzing process and better interpret the results.

\noindent\textbf{$\bullet$ Build Strong Community.}
Establishing a strong community can create a valuable space for users to ask questions, share knowledge, and address encountered issues.

\noindent\textbf{$\bullet$ Implement Replay Features.}
Fuzzer should incorporate exploits replay, as seen in ityFuzz, the only tool supporting replay currently. 
The replay feature can reproduce issues, eliminate false positives, and validate fixes post bug repairs.
}

\section{Threat to validity}
The main threat to external validity in this study is related to the benchmark. 
Firstly, the process of re-labeling contracts involves manual effort, which may introduce potential subjectivity and bias. 
Additionally, in the re-labeling process, some bugs in contracts may not be identified by all fuzzers and thus, could be missed by our volunteers during manual inspection.
Second, the benchmark we used is based on contracts collected from previous works, which may not reflect the diversity and complexity of real-world contracts. 
Therefore, the generalizability of our findings may be constrained.
However, the results can help researchers gain a good understanding of the limitations of existing works and motivate researchers to develop more advanced tools.
In our future work, we will expand our benchmark to include contracts used in production. Additionally, we will use more diverse sources to collect contracts, such as bug bounty programs, security audits.


Another threat to external validity is that, while different fuzzers adopt different techniques, we only evaluated their overall effectiveness, without evaluating the importance of their main components at a more fine-grained level. 
This could lead to a lack of understanding of how the unique features of each fuzzer contribute to their overall performance. In our future work, we will conduct ablation experiments to demonstrate the effects of different unique features of the fuzzers.

Lastly, we recognize that our literature review may not be exhaustive. However, our study represents the most comprehensive investigation of smart contract fuzzers to date. Our future work will systematically compare newly released fuzzers on different blockchain platforms or contract languages.

%% file: Conclusion.tex
\section{Conclusion}
In this paper, we conduct a thorough investigation of existing smart contract fuzzing techniques through a literature review and empirical evaluation. We assess the usability of 11 state-of-the-art smart contract fuzzers using a carefully-labeled benchmark and comprehensive performance metrics. 
Our evaluation identifies some notable issues and suggests possible directions for future fuzzers.
We believe our research offers the community valuable insights, inspiring future breakthroughs in this field.


\section*{Acknowledgment}
We thank the anonymous reviewers for their helpful comments. 
This research is partially supported by the Hong Kong RGC Projects (No. PolyU15219319, PolyU15222320, PolyU15224121), National Natural Science Foundation of China (No. 62272379).